\documentclass[sigplan,10pt]{acmart}

\acmConference[PL'17]{ACM SIGPLAN Conference on Programming Languages}{January 01--03, 2017}{New York, NY, USA}
\acmYear{2017}
\acmISBN{} 
\acmDOI{} 
\startPage{1}

\setcopyright{none}

\usepackage{amsmath}
\usepackage{amssymb}
\usepackage[finalnew]{trackchanges}
\usepackage{stmaryrd}
\usepackage{proof}
\usepackage{graphicx}
\usepackage{listings}
\usepackage{tikz}
\usepackage{gnuplot-lua-tikz}
\usepackage{multicol}
\usepackage{semantic}
\usepackage{setspace}
\usepackage{wrapfig}
\usepackage{color}
\usepackage{caption}
\usepackage{subcaption}
\usepackage{esvect}
\usepackage{tikz}
\usepackage{booktabs} 

\addeditor{Jeremy}
\addeditor{JGS}
\addeditor{Deyaa}
\addeditor{Andre}
\addeditor{AMK}

\newcommand{\code}[1]{\lstinline[basicstyle=\ttfamily,language=Lisp]{#1}}

\definecolor{lightgray}{rgb}{0.95,0.95,0.95}
\definecolor{darkgreen}{rgb}{0.0,0.5,0.0}

\newcommand{\lstsetgrift}{
  \lstset{%
    language=Lisp,
    basicstyle=\ttfamily\small,
    keywordstyle=\ttfamily\bfseries\small,
    columns=flexible,
    aboveskip=\smallskipamount,
    belowskip=\smallskipamount,
    xleftmargin=2pt,
    escapeinside={/+}{+/},
    mathescape=true,
    morekeywords=[1]{define,lambda,if,begin,letrec,let,let*,values,:,repeat,then,else,return,and,time},
    literate=
    {+}{{\textsf{+}}}1
    {-}{{\textsf{-}}}1
    {=>}{{$\rightarrow\;$}}2
    {lambda}{{$\boldsymbol{\lambda}$}}1
  }
}

\def\inline{\lstinline[language=C++,
  basicstyle=\ttfamily\small,
  keywordstyle=\ttfamily\small,
  classoffset=1, keywordstyle={},
  classoffset=2, keywordstyle={},
  classoffset=3, keywordstyle={},
  classoffset=4, keywordstyle={}]}

\newcommand{\castcall}{\inline{cast}}
\newcommand{\coercecall}{\inline{coerce}}
\newcommand{\mkcrcncall}{\inline{mk_crcn}}
\newcommand{\composecall}{\inline{compose}}

\newcommand{\lstsetc}{%
  \lstset{%
    language=C++,
    keywordstyle=\bfseries,
    classoffset=1,
    morekeywords={TAG,UNTAG_INJ,UNTAG_PRJ,UNTAG_SEQ,UNTAG_FUN,
      UNTAG_PREF,UNTAG_MREF,UNTAG_REF,UNTAG_2ND,UNTAG_FUN_PROXY,UNTAG_INJECT,
      UNTAG_FAIL},
    keywordstyle=\color{darkgreen},
    classoffset=2,
    morekeywords={HEAD,TYPE,ARITY,REF_TYPE,UNTAG},
    keywordstyle=\color{blue},
    classoffset=3,
    morekeywords={INJECT,MK_REF_PROXY,MK_FUN_PROXY,caster,MK_SEQ,MK_REF_COERCION},
    keywordstyle=\color{purple},
    classoffset=4,
    morekeywords={cast,coerce,compose,mk_crcn,ref_read,ref_write,compose_fun,},
    keywordstyle=\color{red},
    classoffset=0,
    stringstyle=\color{red},
    commentstyle=\color{darkgreen}%
  }
}

\newcommand{\tinydisplayskips}{%
  \setlength{\abovedisplayskip}{1pt}%
  \setlength{\belowdisplayskip}{5pt}%
  \setlength{\abovedisplayshortskip}{1pt}%
  \setlength{\belowdisplayshortskip}{1pt}}

\newcommand{\newname}[2]{\newcommand{#1}{\ensuremath{#2}}}

\newcommand{\cast}[3]{\ensuremath{#1 \Rightarrow^{#2} #3}}

\newcommand{\IntT}{\ensuremath{\mathsf{Int}}}
\newcommand{\BoolT}{\ensuremath{\mathsf{Bool}}}
\newcommand{\DynT}{\ensuremath{\mathsf{Dyn}}}

\newcommand{\RefPT}[1]{\ensuremath{\mathsf{Ref}_p #1}}
\newcommand{\RefMT}[1]{\ensuremath{\mathsf{Ref}_m #1}}
\newcommand{\FunT}[2]{\ensuremath{#1 \to #2}}

\newcommand{\TupleT}[2]{\ensuremath{#1 \times #2}}

\newcommand{\IntC}{\ensuremath{\mathtt{Int}}}

\newcommand{\DynC}{\ensuremath{\mathtt{Dyn}}}
\newcommand{\RefC}[1]{\ensuremath{\mathtt{Ref}\ #1}}

\newcommand{\idC}{\ensuremath{\mathtt{\iota}}}
\newcommand{\injC}[1]{\ensuremath{#1!}}
\newcommand{\seqInjC}[2]{\ensuremath{\seqC{#1}{\injC{#2}}}}
\newcommand{\prjC}[2]{\ensuremath{#1?^{#2}}}
\newcommand{\seqPrjC}[3]{\ensuremath{\seqC{\prjC{#1}{#2}}{#3}}}
\newcommand{\funC}[2]{\ensuremath{#1 \to #2}}

\newcommand{\refPC}[2]{\ensuremath{\mathsf{Ref}_p\ #1\ #2}}
\newcommand{\refMC}[1]{\ensuremath{\mathsf{Ref}_m\ #1}}
\newcommand{\seqC}[2]{\ensuremath{#1 \, ; \, #2}}
\newcommand{\failC}[3]{\ensuremath{\bot^{#1#2#3}}}
\newcommand{\tupleC}[2]{\ensuremath{#1 \times #2}}

\newcommand{\coercion}[1]{\ensuremath{\langle #1 \rangle}}
\newcommand{\coerce}[2]{\ensuremath{#1 \coercion{#2}}}
\newcommand{\coerced}[2]{\ensuremath{\coerce{#1}{#2}}}
\newcommand{\app}[2]{\ensuremath{#1 \, #2}}
\newcommand{\lam}[2]{\ensuremath{\lambda #1 . #2}}
\newcommand{\blame}[1]{\ensuremath{\mathtt{blame}\, #1}}
\newcommand{\mkrefP}[1]{\ensuremath{\mathtt{ref_P} #1}}
\newcommand{\mkrefM}[1]{\ensuremath{\mathtt{ref_M} #1 @ \AllTa}}
\newcommand{\derefP}[1]{\ensuremath{\mathtt{!_P} #1}}
\newcommand{\derefM}[1]{\ensuremath{\mathtt{!_M} #1}}
\newcommand{\derefTM}[2]{\ensuremath{\mathtt{!} #1 @ #2}}
\newcommand{\setrefP}[2]{\ensuremath{#1 {:=_P} #2}}
\newcommand{\setrefM}[2]{\ensuremath{#1 {:=_M} #2}}
\newcommand{\setrefTM}[3]{\ensuremath{#1 {:=} #2 @ #3}}
\newcommand{\error}{\ensuremath{\mathtt{error}}}

\newcommand{\tuple}[2]{\ensuremath{(#1,#2)}}
\newcommand{\fst}[1]{\ensuremath{(\mathtt{fst}\, #1)}}
\newcommand{\snd}[1]{\ensuremath{(\mathtt{snd}\, #1)}}

\newcommand{\glbt}[2]{\ensuremath{#1 \sqcap #2}}

\newcommand{\var}[1][x]{\ensuremath{#1}}

\newcommand{\const}{\ensuremath{k}}

\newcommand{\uncoerced}{\ensuremath{u}}
\newcommand{\val}{\ensuremath{v}}
\newcommand{\castedval}{\ensuremath{cv}}

\newcommand{\hole}{\ensuremath{\square}}
\newcommand{\context}[2]{\ensuremath{ #1 [ #2 ] }}

\newname{\BaseT}{B}
\newname{\InjTa}{I}
\newname{\InjTb}{J}
\newname{\AllTa}{T}
\newname{\topC}{c}
\newname{\noFailC}{r}
\newname{\blameLabel}{p}
\newname{\addr}{a}
\newname{\seCa}{c}
\newname{\seCb}{d}
\newname{\ca}{c}
\newname{\cb}{d}
\newname{\seFinalC}{i}
\newname{\seInterC}{g}
\newname{\seIdFreeC}{f}
\newname{\CoercionT}{\mathcal{C}}
\newname{\expra}{M}
\newname{\exprb}{N}
\newname{\ctag}{w}
\newname{\uctag}{uc}
\newname{\cctag}{cc}

\newcommand{\composeC}[2]{\ensuremath{#1 \, \fatsemi  \, #2}}
\newcommand{\mkC}[3]{\ensuremath{ #1 \Rightarrow^{#3} #2 }}

\newcommand{\heap}{\ensuremath{\mu}}
\newcommand{\evheap}{\ensuremath{\nu}}
\newcommand{\heapcell}[1]{\ensuremath{\addr \mapsto #1}}
\newcommand{\mheapcell}[2]{\ensuremath{\addr \mapsto #1 : #2}}

\newcommand{\LowMonotonicVsRacket}{\ensuremath{0.42}}
\newcommand{\HighMonotonicVsRacket}{\ensuremath{2.36}}

\newcommand{\MonotonicVsProxiedMatmult}{\ensuremath{1.48}}
\newcommand{\MonotonicVsProxiedNbody}{\ensuremath{1.08}}

\newcommand{\MonotonicVsProxiedOnPartial}{\ensuremath{1.48}}

\acmDOI{0000001.0000001}

\received{November 2017}

\begin{document}
\title{Efficient Gradual Typing}


\author{Andre Kuhlenschmidt}
 \affiliation{
   \institution{Indiana University}            
}
\email{akuhlens@indiana.edu}          

\author{Deyaaeldeen Almahallawi}
\affiliation{\institution{Indiana University}}
\email{dalmahal@indiana.edu}

\author{Jeremy G. Siek}
\affiliation{\institution{Indiana University}}
\email{jsiek@indiana.edu}



\begin{abstract}
  Gradual typing combines static and dynamic typing in the same
  program.  One would hope that the performance in a gradually typed
  language would range between that of a dynamically typed language
  and a statically typed language.  Existing implementations of
  gradually typed languages have not achieved this goal due to
  overheads associated with runtime casts. Takikawa et al. (2016)
  report up to $100\times$ slowdowns for partially typed programs.  In
  this paper we present a compiler, named Grift, for evaluating
  implementation techniques for gradual typing. We take a
  straightforward but surprisingly unexplored implementation approach
  for gradual typing, that is, ahead-of-time compilation to native
  assembly code with carefully chosen runtime representations
  and space-efficient coercions.

  Our experiments show that this approach achieves performance on par
  with OCaml on statically typed programs and performance between that
  of Gambit and Racket on untyped programs.
  On partially typed code, the geometric mean ranges from
  $\LowMonotonicVsRacket\times$ to $\HighMonotonicVsRacket\times$ that
  of (untyped) Racket across the benchmarks.
  We implement casts using the coercions of Siek, Thiemann, and Wadler
  (2015). This technique eliminates all catastrophic slowdowns without
  introducing significant overhead. Across the benchmarks, coercions
  range from 15\% slower (fft) to almost 2$\times$ faster (matmult)
  than regular casts.
  We also implement the monotonic references of Siek et al. (2015).
  Monotonic references eliminate all overhead in statically typed
  code, and for partially typed code, they are faster than proxied
  references, sometimes up to $\MonotonicVsProxiedOnPartial\times$.
%
\end{abstract}




\maketitle

\section{Introduction}

Gradual typing combines static and dynamic type checking in the same
program, giving the programmer control over which typing discipline is
used for each region of
code~\cite{Anderson:2002kd,Siek:2006bh,Tobin-Hochstadt:2006fk,Gronski:2006uq,Matthews:2007zr}. We
would like gradually typed languages to be \emph{efficient},
\emph{sound}, and provide \emph{interoperability}. Regarding
efficiency, we would like the performance of gradual typing to range
from being similar to that of a dynamically typed language to that of
a statically typed language.  Regarding soundness, programmers (and
compilers) would like to trust type annotations and know that runtime
values respect their compile-time types. Third, regarding
interoperability, static and dynamic regions of code should
interoperate seamlessly.

To date, implementations of gradual typing have only delivered
two of these three properties.  For example, Typed
Racket~\cite{Tobin-Hochstadt:2008lr} provides soundness and
interoperability but suffers from slowdowns of up to
$100\times$~\cite{Takikawa:2015aa,Takikawa:2016aa} on a partially
typed program. Thorn~\cite{Bloom:2009aa,Wrigstad:2010fk} and Safe
TypeScript~\cite{Rastogi:2014aa} provide better performance but limit
interoperability. TypeScript~\cite{Hejlsberg:2012aa,Bierman:2014aa}
and Gradualtalk~\cite{Allende:2011fk,Allende:2012aa,Allende:2013aa} do
not provide soundness and their performance is on par with dynamic
languages but not static ones, but they provide seamless
interoperability.


Several papers at OOPSLA 2017 begin to address the efficiency concerns
for gradually typed languages that are committed to soundness and
interoperability.  \citet{Bauman:2017aa} demonstrate that a tracing
JIT can eliminate 90\% of the overheads in Typed Racket due to gradual
typing. \citet{Richards:2017aa} augment the Higgs JIT compiler and
virtual machine (VM)~\citep{Chevalier-Boisvert:2015aa} for JavaScript,
re-purposing the VM's notion of \emph{shape} to implement monotonic
references~\citep{Siek:2015aa}. \citet{Richards:2017aa} reports that
this reduces the worst slowdowns to 45\%, with an average slowdown of
just 7\%. Meanwhile, \citet{Muehlboeck:2017aa} show that for
nominally-typed object-oriented languages, efficiency is less of a
problem.

In this paper we demonstrate that efficient gradual typing can be
achieved in structurally-typed languages by relatively straightforward
means. We build and evaluate an ahead-of-time compiler that uses
carefully chosen runtime representations and implements two important
ideas from the theory of gradual typing.  It uses space efficient
\emph{coercions}~\citep{Siek:2009ys,Herman:2010aa,Garcia:2013fk,Siek:2015ab}
to implement casts and it reduces overhead in statically typed code by
using \emph{monotonic references}~\citep{Siek:2015aa}.

\paragraph{Contributions} This paper makes these contributions.

\begin{itemize}
\item A space-efficient semantics for monotonic references and lazy-D
  coercions (Section~\ref{sec:semantics}).
  
\item The first ahead-of-time compiler, named Grift, for a gradually
  typed language that targets native assembly code.  The compiler is
  the first to implement space-efficient coercions
  (Section~\ref{sec:implementation}).

\item Experiments (Section~\ref{sec:external-comparison}) showing 
  \begin{itemize}
  \item performance on statically typed code that is
    on par with OCaml,
  \item performance on dynamically typed code  that 
    is between Gambit and
    Racket, and
  \item performance on partially typed code ranging from
    $\LowMonotonicVsRacket\times$ to $\HighMonotonicVsRacket\times$
    that of Racket.
  \end{itemize}
  
\item Experiments showing that coercions eliminate catastrophic
  slowdowns without adding significant overhead
  (Section~\ref{sec:cost-space-efficiency}).

\item Experiments showing that monotonic references eliminate overhead
  in statically typed code
  (Section~\ref{sec:monotonic-versus-proxied}).

\end{itemize}

Section~\ref{sec:gradual-typing} provides background on gradual
typing, focusing on runtime casts and the tension between efficiency,
soundness, and interoperability.

\section{Tensions in Gradual Typing}
\label{sec:gradual-typing}

From a language design perspective, gradual typing touches both the
type system and the operational semantics.  The key to the type system
is the \emph{consistency} relation on types, which enables implicit
casts to and from the unknown type, here written \DynC{}, while still
catching static type
errors~\cite{Anderson:2002kd,Siek:2006bh,Gronski:2006uq}.  The dynamic
semantics for gradual typing is based on the semantics of
\emph{contracts}~\cite{Findler:2002es,Gray:2005ij},
\emph{coercions}~\cite{Henglein:1994nz}, and \emph{interlanguage
migration}~\cite{Tobin-Hochstadt:2006fk,Matthews:2007zr}. Because of
the shared mechanisms with these other lines of research, much of the
ongoing research in those areas benefits the theory of gradual typing,
and vice
versa~\cite{Guha:2007kl,Matthews:2008qr,Greenberg:2010lq,Dimoulas:2011fk,Strickland:2012fk,Chitil:2012aa,Dimoulas:2012fk,Greenberg:2015ab}.
In the following we give a brief introduction to gradual typing by way
of an example that emphasizes the three main goals of gradual typing:
supporting interoperability, soundness, and efficiency.

\paragraph{Interoperability and Evolution}

Consider the example program in Figure~\ref{fig:egcd}, written in a
variant of Typed Racket that we have extended to support fine-grained
gradual typing.  On the left side of the figure we have an untyped
function for the extended greatest common divisor. With gradual
typing, unannotated parameters are dynamically typed and therefore
assigned the type \DynC{}. On the right side of the figure is the same
function at a later point in time in which the parameter types have
been specified (\IntC{}) but not the return type. With gradual typing,
both programs are well typed because implicit casts are allowed to and
from \DynC{}. For example, on the left we have the expression
\code{(modulo b a)}, so \code{b} and \code{a} are implicitly cast from
\DynC{} to \IntC{}. On the right, there is an implicit cast around
\code{(list b 0 1)} from \code{(List Int)} to \DynC{}.  The reason
that gradual typing allows implicit casts both to and from \DynC{} is
to enable evolution. As a programmer adds or removes type annotations,
the program continues to type check and also exhibits the same
behavior up to cast errors, a property called the \emph{gradual
  guarantee}~\cite{Siek:2015ac}.

\begin{figure*}[tbp]
\lstsetgrift
\begin{tabular}{cc}
\begin{minipage}{0.5\textwidth}
\begin{lstlisting}
(define (egcd a b)
  (if (= a 0) 
    (list b 0 1)
    (let ([r (egcd (modulo b a) a)])
      (list (car r)
        (- (caddr r) (* (/ b a) (cadr r)))
        (cadr r)))))
\end{lstlisting}
\end{minipage}
&
\begin{minipage}{0.5\textwidth}
\begin{lstlisting}
(define (egcd [a:Int] [b:Int])
  (if (= a 0) 
    (list b 0 1)
    (let ([r (egcd (modulo b a) a)])
      (list (car r)
        (- (caddr r) (* (/ b a) (cadr r)))
        (cadr r)))))
\end{lstlisting}
\end{minipage}
\end{tabular}
\caption{Two gradually typed versions of extended GCD.}
\label{fig:egcd}
\end{figure*}

\paragraph{Soundness}

Next consider the function \code{modinv} defined below that computes
the modular inverse using the second version of \code{egcd}.  What
happens when the code on the right forgets to convert the input string
from \code{(read)} to an integer before passing it to \code{modinv}?
%

\lstsetgrift
\hspace{-0.6cm}
\begin{minipage}[t]{0.55\linewidth}
\begin{lstlisting}
(define (modinv a m)
  (let ([r (egcd a m)])
    (if (not (= (car r) 1))
        (error ...)
        (modulo (cadr r) m))))
\end{lstlisting}
\end{minipage}
\begin{minipage}[t]{0.45\linewidth}
\begin{lstlisting}
(let ([input (read)])
  (modinv 42 input))
\end{lstlisting}
\end{minipage}
%

%
Parameter \code{m} of \code{modinv} has type \DynC{}, but \code{b} of
\code{egcd} has type \code{Int}, so there is an implicit cast from
\DynC{} to \code{Int}. With gradual typing, this implicit cast comes
with a runtime cast that will trigger an error if the input to this
program is a string. This runtime cast is required to ensure
soundness: without it a string could flow into \code{egcd} and
masquerade as an \code{Int}. Soundness is not only important for
software engineering reasons but it also impacts efficiency both
positively and negatively.

Ensuring soundness in the presence of first-class functions and
mutable references is nontrivial. When a function is cast from \DynC{}
to a type such as $\IntC \to \IntC$, it is not possible for the cast
to know whether the function will return an integer on all
inputs. Instead, the standard approach is to wrap the function in a
proxy that checks the return value each time the function is
called~\citep{Findler:2002es}. Similarly, when a mutable reference is
cast, e.g., from $\RefC{\IntC}$ to $\RefC{\DynC}$, the reference is
wrapped in a proxy that casts from $\IntC$ to $\DynC$ on every read
and from $\DynC$ to $\IntC$ on every
write~\citep{Herman:2006uq,Herman:2010aa}.

\paragraph{Efficiency}

Ideally, statically typed code within a gradually typed program should
execute without overhead. Likewise, partially typed or untyped code
should execute with no more overhead than is typical of dynamically
typed languages. Consider the \code{egcd} on the right side of
Figure~\ref{fig:egcd}. Inside this \code{egcd}, the expression
\code{(modulo b a)} should simply compile to an \code{idiv}
instruction (on x86).  However, if the language did not ensure
soundness as discussed above, then this efficient compilation strategy
would result in undefined behavior (segmentation faults at best,
hacked systems at worst). It is soundness that enables type-based
specialization. However, soundness comes at the cost of the runtime
casts at the boundaries of static and dynamic code.

\section{Semantics of a Gradual Language}
\label{sec:semantics}

The type system of Grift's input language is the standard one for the
gradually typed lambda
calculus~\citep{Siek:2006bh,Siek:2008aa,Herman:2010aa}.  The
operational semantics, as usual, is expressed by a translation to an
intermediate language with explicit casts.

\begin{figure*}[tbp]
\lstsetgrift %
\begin{tabular}{cc}
\begin{minipage}{0.4\textwidth}
  Source Program:
  \begin{lstlisting}
(let ([add1 : (Int => Int)
       (lambda ([x : Int]) (+ x 1))])
  (let ([f : (Dyn => Dyn) add1])
    (: (f 41) Int)))
  \end{lstlisting}
\end{minipage}
&
\begin{minipage}{0.6\textwidth}
  After Cast Insertion:
  \begin{lstlisting}
(let ([add1 (lambda (x) (+ x 1))])
  (let ([f (cast add1 (/+\FunT{\IntT}{\IntT}+/) (/+\FunT{\DynT}{\DynT}+/) L1)])
    (cast (f (cast 41 /+\IntT+/ /+\DynT+/ L2)) /+\DynT+/ /+\IntT+/ L3)))
  \end{lstlisting} 
\end{minipage} \\
\end{tabular}
\caption{An example of the Grift compiler inserting casts.  The \lstinline{L1},
    \lstinline{L2}, etc. are blame labels that identify source code location.}
  \label{fig:schml-cast-insertion}
\end{figure*}

\lstsetgrift
Consider the source program in Figure~\ref{fig:schml-cast-insertion}
which calculates the value 42 by applying the \lstinline{add1}
function, by way of variable \lstinline{f}, to the integer value
41. The type of \lstinline{add1} does not exactly match the type
annotation on \lstinline{f} (which is \lstinline{Dyn => Dyn}) so the
compiler inserts the cast:

\centerline{ \lstinline{(cast add1 (Int => Int) (Dyn  => Dyn) l2)}}

\noindent
The application of \lstinline{f} to \lstinline{42} requires a cast on
\lstinline{42} from \lstinline{Int} to \lstinline{Dyn}.  Also, the
return type of \lstinline{f} is \lstinline{Dyn}, so the compiler 
inserts a cast to convert the returned value to \lstinline{Dyn} to
satisfy the type ascription.

In this paper we consider two approaches to the implementation of
runtime casts: traditional casts, which we refer to as
\emph{type-based casts}, and \emph{coercions}.  Type-based casts
provide the most straightforward implementation, but the proxies they
generate can accumulate and consume an unbounded amount of
space~\citep{Herman:2010aa}. The coercions of \citet{Henglein:1994nz}
solve the space problem with a representation that enables the
compression of higher-order casts~\citep{Herman:2010aa}.

For type-based casts, the dynamic semantics that we use is almost
covered in the literature.
We use the lazy-D cast semantics which is described by
\citet{Siek:2012uq}. (They were originally described using coercions by
\citet{Siek:2009rt}.) The distinction between lazy-D and the more
commonly used lazy-UD semantics \citep{Wadler:2009qv} is not well-known,
so to summarize the difference: in lazy-D, arbitrary types of values may
be directly injected into type $\DynT$, whereas in lazy-UD, only values
of a \emph{ground} type may be directly injected into $\DynT$. 
For example, $\IntT$ and $\DynT \to \DynT$ are ground types,
but $\IntT \to \IntT$ is not.


The one missing piece for our purposes are the reduction rules for
proxied references, which we adapt from the coercion-based version by
\citet{Herman:2010aa}.  In this setting, proxied references are values
of the form $(v : \RefPT{T_1} \Rightarrow^\ell \RefPT{T_2})$. The
following are the reduction rules for reading and writing to a proxied
reference.
\tinydisplayskips
\begin{align*}
  !(v : \RefPT{T_1} \Rightarrow^\ell \RefPT{T_2})
  & \longrightarrow \; !v : T_1 \Rightarrow^\ell T_2 \\
  (v_1 : \RefPT{T_1} \Rightarrow^\ell \RefPT{T_2}) := v_2
  & \longrightarrow v_1 := (v_2 : T_2 \Rightarrow^\ell T_1)
\end{align*}
For monotonic references with type-based casts, the dynamic semantics
for lazy-D is given by \citet{Siek:2015aa}.

Regarding coercions, the dynamic semantics that we used is less
well-covered in the literature. Again, we use the lazy-D semantics of
\citet{Siek:2009rt}, but that work, despite using coercions, did not
define a space-efficient semantics. On the other hand,
\citet{Siek:2015ab} give a space-efficient semantics with coercions, but
for the lazy-UD semantics. To this end, they define a normal form for
coercions and a composition operator that compresses coercions.  Here we
adapt that approach to lazy-D, which requires some changes to the normal
forms and to the composition operator. Also, that work did not consider
mutable references, so here we add support for both proxied and
monotonic references. Regarding monotonic references,
\citet{Siek:2015aa} define the lazy-D semantics, but again, they did not
define a space-efficient semantics. Here we make it space-efficient by
defining the normal forms for reference coercions and the composition
operation on them.

\begin{figure}[tbp]
  \flushleft
  Types and coercions
  \[  
    \begin{array}{llcl}
      \text{Base Types}
      & \BaseT  & ::= & \IntT \mid \BoolT \mid \ldots \\
      \text{Injectables}
      & \InjTa,\InjTb  & ::= &
                               \BaseT \mid
                               \FunT{\AllTa}{\AllTa} \mid
                             \TupleT{\AllTa}{\AllTa} \mid
                             \RefPT{\AllTa} \mid
                             \RefMT{\AllTa}\\
      \text{Types}
      & \AllTa  & ::= &
                      \DynT \mid
                      \InjTa\\              
      
      \text{Coercions}
      & \seCa,\seCb & ::= &
                          \idC \mid
                          (\seqPrjC{\InjTa}{\blameLabel}{\seFinalC}) \mid
                          \seFinalC \\
      \text{Final Crcns}
      & \seFinalC & ::= &
                        \failC{\InjTa}{\blameLabel}{\InjTb}        \mid
                        (\seqInjC{\seInterC}{\InjTa}) \mid
                        \seInterC                  \\
      \text{Mid. Crcns}
      & \seInterC & ::= &
                        \idC                  \mid
                        \funC{\seCa}{\seCb}   \mid
                        \tupleC{\seCa}{\seCb} \mid
                        \refPC{\seCa}{\seCb}  \mid
                        \refMC{\AllTa} \\
      \text{Id-free Crcns}
      & \seIdFreeC & ::= & (\seqPrjC{\InjTa}{\blameLabel}{\seFinalC})\mid
                         (\seqInjC{\seInterC}{\InjTa})\mid
                         \funC{\seCa}{\seCb}\mid
                         \tupleC{\seCa}{\seCb} \\
                         &&& \refPC{\seCa}{\seCb}\mid
                         \refMC{\AllTa}\mid
                         \failC{\InjTa}{\blameLabel}{\InjTb}
    \end{array}
    \]
Consistency \hfill \fbox{$T \sim T$}
\begin{gather*}
\DynT \sim T
\qquad
T \sim \DynT
\qquad
\BaseT \sim \BaseT
\\
\inference
  {T_1 \sim T_2}
  {\RefPT{T_1} \sim \RefPT{T_2}}
\quad
\inference
  {T_1 \sim T_2}
  {\RefMT{T_1} \sim \RefMT{T_2}}
\\[1ex]
\inference
  {T_1 \sim T_3 & T_2 \sim T_4}
  {T_1 \to T_2 \sim T_3 \to T_4}
\quad
\inference
  {T_1 \sim T_3 & T_2 \sim T_4}
  {T_1 \times T_2 \sim T_3 \times T_4}
\end{gather*}
Meet operation (greatest lower bound) \hfill \fbox{$T \sqcap T$} 
\begin{align*}
\DynT \sqcap T = T \sqcap \DynT &= T \\
\BaseT \sqcap \BaseT &= \BaseT\\
(T_1 \times T_2) \sqcap (T_3 \times T_4) &=
  (T_1 \sqcap T_3) \times (T_2 \sqcap T_4) \\
(T_1 \to T_2) \sqcap (T_3 \to T_4) &=
  (T_1 \sqcap T_3) \to (T_2 \sqcap T_4) \\
\RefPT{T_1} \sqcap \RefPT{T_2} &= \RefPT{(T_1 \sqcap T_2)}\\
\RefMT{T_1} \sqcap \RefMT{T_2} &= \RefMT{(T_1 \sqcap T_2)}
\end{align*} \\
Coercion creation \hfill \fbox{$(\cast{T}{l}{T}) = c$} 
\begin{align*}
  (\cast{\BaseT}{l}{\BaseT}) = (\cast{\DynT}{l}{\DynT})
  & = \iota \\
  (\cast{\InjTa}{l}{\DynT})
  & = \injC{\InjTa} \\
  (\cast{\DynT}{l}{\InjTa})
  & = \prjC{\InjTa}{l} \\
  (\cast{\FunT{\AllTa_1}{\AllTa_2}}{l}{\FunT{\AllTa_1'}{\AllTa_2'}})
  & =
    \funC{(\cast{\AllTa_1'}{l}{\AllTa_1})}{(\cast{\AllTa_2}{l}{\AllTa_2'})}\\
  (\cast{\TupleT{\AllTa_1}{\AllTa_2}}{l}{\TupleT{\AllTa_1'}{\AllTa_2'}})
  & =
    \tupleC{(\cast{\AllTa_1}{l}{\AllTa_1'})}{(\cast{\AllTa_2}{l}{\AllTa_2'})} \\
  (\cast{\RefPT{\AllTa}}{l}{\RefPT{\AllTa'}})
  & =
    \refPC{(\cast{\AllTa'}{l}{\AllTa})}{(\cast{\AllTa}{l}{\AllTa'})}\\
 (\cast{\RefMT{\AllTa}}{l}{\RefMT{\AllTa'}})
  & =
    \refMC{\AllTa'} \\
\end{align*}

Coercion composition \hfill \fbox{$\composeC{\seCa}{\seCb} = r$}\\
\begin{minipage}{0.45\textwidth}
  \[
    \begin{array}{rcl}
            \composeC{(\seqC{\seInterC}{\injC{\InjTa}})}
            {(\seqPrjC{\InjTb}{\blameLabel}{\seFinalC})}
      & = &
            \composeC{\seInterC}
            ({\composeC{\langle\mkC{\InjTa}{\InjTb}{\blameLabel}\rangle}
            {\seFinalC}}) \\
      
      \composeC{\funC{\seCa}{\seCb}}
      {\funC{\seCa'}{\seCb'}} 
                                        & = &
                                              \funC{(\composeC{\seCa'}{ \seCa})}
                                              {(\composeC{\seCb}{
                                              \seCb'})}\\

      \composeC{\tupleC{\seCa}{\seCb}}
      {\tupleC{\seCa'}{\seCb'}} 
                                        & = &
                                              \tupleC{(\composeC{\seCa}{ \seCa'})}
                                              {(\composeC{\seCb}{ \seCb'})}\\
      
      \composeC{\refPC{\seCa}{\seCb}}
      {\refPC{\seCa'}{\seCb'}}
                                        & = &
                                              \refPC{(\composeC{\seCa'}{\seCa})}
                                              {(\composeC{\seCb}{\seCb'})}\\
      \composeC{\refMC{\AllTa}}
      {\refMC{\AllTa'}}
                                        & = &
      \refMC{(\glbt{\AllTa}{\AllTa'})}
    \end{array}
    \]
\end{minipage}
\begin{minipage}{0.45\textwidth}
  \[
      \begin{array}{rcl}      
           \composeC{(\seqPrjC{\InjTa}{\blameLabel}{\seFinalC})}
                                                           {\seCa}
                                                       & = &
                                                             \seqPrjC{\InjTa}{\blameLabel}{(\composeC{\seFinalC}{\seCa})}\\

      \composeC{\seInterC_1}{(\seqInjC{\seInterC_2}{\InjTa})}
                                        & = &  
                                              \seqInjC{(\composeC{\seInterC_1}{\seInterC_2})}{\InjTa} \\
      
      \composeC{\idC}{\seCa} =
      \composeC{\seCa}{\idC} & = & \seCa\\

      \composeC{\seInterC}{\failC{\InjTa}{\blameLabel}{\InjTb}} =
      \composeC{\failC{\InjTa}{\blameLabel}{\InjTb}}{\seCa}      & = &   \failC{\InjTa}{\blameLabel}{\InjTb} \\
    \end{array}
  \]
\end{minipage}
\caption{Types, coercions, and their operations.}
\label{fig:coercions}
\end{figure}

Figure~\ref{fig:coercions} defines a representative subset of the types
and coercions used in Grift's intermediate language.  The figure also
defines the \emph{meet} operation and the \emph{consistency} relation on
types and the \emph{composition} operator on coercions.  Instead of
defining two different languages, one with proxied references and the
other with monotonic references, we instead present a single language
with both kinds of references. The type $\RefPT{T}$ is for proxied
references and $\RefMT{T}$ is for monotonic references.  Likewise,
$\refPC{\seCa}{\seCb}$ is the coercion for proxied references and
$\refMC{T}$ is the coercion for monotonic references.

Space-efficient coercions are defined by a grammar consisting of three
rules that enable coercion composition by the composition operator
defined in Figure~\ref{fig:coercions}. Let $\seCa,\seCb$ range over
space-efficient coercions, $\seFinalC$ range over final coercions, and
$\seInterC$ range over middle coercions. Space-efficient coercions are
either the identity coercion $\idC$, a projection followed by a final
coercion $(\seqPrjC{\InjTa}{\blameLabel}{\seFinalC})$, or just a final
coercion. A final coercion is either the failure coercion
$\failC{\InjTa}{\blameLabel}{\InjTb}$, a middle coercion followed by
an injection $(\seqInjC{\seInterC}{\InjTa})$, or just an intermediate
coercion. An intermediate coercion is either the identity coercion
$\idC$, the function coercion $\funC{\seCa}{\seCb}$, the tuple
coercion $\tupleC{\seCa}{\seCb}$, the proxied reference coercion
$\refPC{\seCa}{\seCb}$, where $\seCa$ is applied when writing and
$\seCb$ is applied when reading, or the monotonic reference coercion
$\refMC{\AllTa}$. The main difference between the lazy-D coercions
shown here and those of \citet{Siek:2015ab} is in the injection $I!$
and projection $J?^p$ coercions, which take any \emph{injectable} type
(anything but $\DynT$) instead of only ground types.  This change
impacts the coercion composition operation, in the case where an
injection $I!$ meets a projection $J?^p$ we make a new coercions whose
source is $I$ and target is $J$ with the coercion creation operation
$\langle\mkC{\InjTa}{\InjTb}{\blameLabel}\rangle$
(Figure~\ref{fig:coercions}).

\begin{figure}[tbp]
  \flushleft
  Runtime Structures
  \[  
    \begin{array}{lcl}
       \castedval & ::= & \val \mid \coerced{\val}{\seCa} \mid \tuple{\castedval}{\castedval} \\
       \heap & ::= & \emptyset \mid \heap(\heapcell{v}) \mid
                    \heap(\mheapcell{v}{\AllTa}) \\
       \evheap & ::= & \heap \mid \evheap(\mheapcell{\castedval}{\AllTa})\\
      \mathcal{E} & ::= & \mathcal{F} \mid
                            \context{\mathcal{F}}{\coerce{\hole}{\seIdFreeC}}\\
      \mathcal{F} & ::= & 
                            \hole \mid
                            \context{\mathcal{E}}{\app{\hole}{\expra}} \mid
                            \context{\mathcal{E}}{\app{v}{\hole}} \mid
                            \context{\mathcal{E}}{\tuple{\hole}{\expra}} \mid
                            \context{\mathcal{E}}{\tuple{v}{\hole}} \mid \\
                     & & 
                            \context{\mathcal{E}}{\fst{\hole}} \mid
                            \context{\mathcal{E}}{\snd{\hole}} \mid
                           \context{\mathcal{E}}{\mkrefP{\hole}} \mid
                           \context{\mathcal{E}}{\derefP{\hole}} \mid \\
                     & & 
                           \context{\mathcal{E}}{\setrefP{\hole}{\expra}} \mid
                           \context{\mathcal{E}}{\setrefP{v}{\hole}} \mid
                           \context{\mathcal{E}}{\mkrefM{\hole}} \mid
                           \context{\mathcal{E}}{\derefM{\hole}} \mid\\
                     & & 
                           \context{\mathcal{E}}{\derefTM{\hole}{\AllTa}} \mid
                           \context{\mathcal{E}}{\setrefM{\hole}{\expra}} \mid
                           \context{\mathcal{E}}{\setrefM{v}{\hole}} \mid \\
                     & & 
                           \context{\mathcal{E}}{\setrefTM{\hole}{\expra}{\AllTa}} \mid
                           \context{\mathcal{E}}{\setrefTM{v}{\hole}{\AllTa}}
    \end{array}
  \]
  
  \text{Cast reduction rules}\hfill \fbox{$\expra,\heap \longrightarrow_{c} \exprb,\evheap$} \\
  \[
    \begin{array}{rcl}
      \context{\mathcal{F}}{\coerce{\uncoerced}{\idC}},\heap
      & \longrightarrow_{c} &
                          \context{\mathcal{F}}{\uncoerced},\heap\\
      \context{\mathcal{F}}{\coerce{\coerced{\uncoerced}{\seFinalC}}{\seCa}},\heap
      & \longrightarrow_{c} &
                          \context{\mathcal{F}}{\coerced{\uncoerced}{\composeC{\seFinalC}{\seCa}}},\heap\\
      \context{\mathcal{F}}{\coerced{\tuple{\uncoerced}{\uncoerced'}}{\tupleC{\seCa}{\seCb}}},\heap
      & \longrightarrow_{c} & 
                          \context{\mathcal{E}}{\tuple{\coerced{\uncoerced}{\seCa}}{\coerced{\uncoerced'}{\seCb}}},
                              \heap \\
      \context{\mathcal{F}}{\coerce{\uncoerced}{\failC{\InjTa}{\blameLabel}{\InjTb}}},\heap
      & \longrightarrow_{c} &
                          \blame{\blameLabel},\heap \\
     \context{\mathcal{F}}{\coerced{\addr}{\refMC{\AllTa_2}}},\evheap 
     & \longrightarrow_{c} & 
     \context{\mathcal{F}}{a},\evheap(\mheapcell{\coerce{\castedval}{\cast{\AllTa_1}{}{\AllTa_3}}}{\AllTa_3})\\
     && \text{if } \evheap(a)=\castedval {:} \AllTa_1, \AllTa_3{=}\glbt{\AllTa_1}{\AllTa_2}, \AllTa_3 \neq \AllTa_1 \\
     \context{\mathcal{F}}{\coerced{\addr}{\refMC{\AllTa_2}}},\evheap
     &\longrightarrow_{c} &\context{\mathcal{F}}{a},\evheap\\
     &&\text{if } \evheap(a)=\castedval : \AllTa_1,
        \AllTa_1=\glbt{\AllTa_1}{\AllTa_2} \\
     \context{\mathcal{F}}{\coerced{\addr}{\refMC{\AllTa_2}}},\evheap
     &\longrightarrow_{c}& \error,\evheap \\
     &&\text{if } \evheap(a)=\castedval : \AllTa_1 \nsim \AllTa_2
    \end{array}
  \]
   \\
  \text{Program reduction rules}\hfill \fbox{$\expra,\heap \longrightarrow_{e} \exprb,\evheap$} \\
\begin{minipage}{0.42\textwidth}
\small
  \[
    \begin{array}{rcl}
      \context{\mathcal{E}}{(\lam{x}{\expra})\ \val},\heap
      & \longrightarrow_{e} & \context{\mathcal{E}}{[x:=\val]\expra},\heap\\
      \context{\mathcal{E}}{\app{\coerced{\uncoerced}{\funC{\seCa}{\seCb}}}{\val}},\heap
      & \longrightarrow_{e} & 
                          \context{\mathcal{E}}{\coerce{\app{\uncoerced}{(\coerce{\val}{\seCa})}}{\seCb}},
                              \heap \\
      \context{\mathcal{E}}{\fst{\tuple{\val}{\val'}}},\heap
      & \longrightarrow_{e} & \context{\mathcal{F}}{\val},\heap\\
      \context{\mathcal{E}}{\snd{\tuple{\val}{\val'}}},\heap
      & \longrightarrow_{e} & \context{\mathcal{F}}{\val'},\heap\\
      \context{\mathcal{E}}{\blame{\blameLabel}},\heap
      & \longrightarrow_{e} & \blame{\blameLabel},\heap \quad \text{if }
                                                         \mathcal{E} \neq \hole\\
      \context{\mathcal{E}}{\error},\heap & \longrightarrow_{e} &
      \error,\heap
      \quad \text{if } \mathcal{E} \neq \hole \\
     \context{\mathcal{E}}{\mkrefP{\val}},\heap
      & \longrightarrow_{e} &
                          \context{\mathcal{F}}{\addr},\heap(\heapcell{\val}) \quad \text{if }\addr \not\in \text{dom}(\heap)\\
      \context{\mathcal{E}}{\derefP{\addr}},\heap
      & \longrightarrow_{e} & \context{\mathcal{F}}{\heap(\addr)},\heap\\
      \context{\mathcal{E}}{\derefP{(\coerced{\addr}{\refPC{\seCa}{\seCb}})}},\heap
      & \longrightarrow_{e} &
                          \context{\mathcal{F}}{\coerce{(\derefP{\addr})}{\seCb}},\heap\\
      \context{\mathcal{E}}{\setrefP{\addr}{\val}},\heap
      & \longrightarrow_{e} & \context{\mathcal{F}}{\addr},\heap(\heapcell{\val})\\
     \context{\mathcal{E}}{\setrefP{\coerced{\addr}{\refPC{\seCa}{\seCb}}}{\val}},\heap
      & \longrightarrow_{e} &
                          \context{\mathcal{E}}{\setrefP{\addr}{\coerce{\val}{\seCa}}},\heap
     \\
      \context{\mathcal{E}}{\mkrefM{\val}},\heap
      & \longrightarrow_{e} &
                          \context{\mathcal{F}}{\addr},\heap(\mheapcell{\val}{\AllTa})\quad\text{if }\addr \not\in\text{dom}(\heap)\\
      \context{\mathcal{E}}{\derefM{\addr}},\heap
      & \longrightarrow_{e} & \context{\mathcal{F}}{\heap(\addr)_{\text{val}}},\heap\\
      \context{\mathcal{E}}{\derefTM{\addr}{\AllTa}},\heap
      & \longrightarrow_{e} & \context{\mathcal{F}}{\coerced{\heap(\addr)_{\text{val}}}{\cast{\heap(\addr)_{\text{rtti}}}{}{\AllTa}}},\heap\\
      \context{\mathcal{E}}{\setrefM{\addr}{\val}},\heap
      & \longrightarrow_{e} & \context{\mathcal{F}}{\addr},\heap(\mheapcell{\val}{\heap(\addr)_{\text{rtti}}})\\
      \context{\mathcal{E}}{\setrefTM{\addr}{\val}{\AllTa}},\heap
      & \longrightarrow_{e} &
                          \context{\mathcal{F}}{\addr},\heap(\mheapcell{\val\langle\cast{\AllTa}{}{\heap(\addr)_{\text{rtti}}\rangle}}{\heap(\addr)_{\text{rtti}}}) 
    \end{array}
  \]
\end{minipage}  \\[1ex]
  \text{State reduction rules}\hfill \fbox{$\expra,\evheap
    \longrightarrow \exprb,\evheap$}
  \begin{gather*}
    \inference
    {\expra,\heap\longrightarrow_{X}\exprb,\evheap \\
      X\in\{c,e\}}
    {\expra,\heap\longrightarrow\exprb,\evheap} 
    \quad
    \inference
    {\evheap(\addr)=\castedval : \AllTa \quad
     \evheap'(\addr)_{\text{rtti}}=\AllTa\\
    \castedval,\evheap\longrightarrow_{c}\castedval',\evheap'
    }
    {\expra,\evheap\longrightarrow\expra,\evheap'(\mheapcell{\castedval'}{\AllTa})}
    \\
    \inference
    {\evheap(\addr)=\castedval : \AllTa \;
    \evheap'(\addr)_{\text{rtti}}\neq\AllTa\\
    \castedval,\evheap\longrightarrow_{c}\castedval',\evheap'
    }
    {\expra,\evheap\longrightarrow\expra,\evheap'}
    \;
    \inference
    {\evheap(\addr)=\castedval : \AllTa \\
     E \in\{\error,\blame{p}\} \\
     \castedval,\evheap\longrightarrow_{c}E,\evheap'}
    {\expra,\evheap\longrightarrow E,\evheap'}
  \end{gather*}
  
  \caption{Semantics of the intermediate language of Grift, with
    proxied and monotonic references.}
  \label{fig:semantics}
\end{figure}

The following is the syntax of Grift's intermediate language,
including both proxied and monotonic references.
\begin{align*} 
       \uncoerced
                     & ::= 
                             \const \mid
                             \addr \mid
                             \lam{x}{\expra} \mid
                             \tuple{\uncoerced}{\uncoerced}\\
       \val & ::= 
                     \uncoerced \mid
                     \tuple{\val}{\val} \mid
                   \coerced{\uncoerced}{\seqInjC{\seInterC}{\InjTa}} \mid
                   \coerced{\uncoerced}{\funC{\seCa}{\seCb}} \mid
                   \coerced{\uncoerced}{\refPC{\seCa}{\seCb}} \\
       b & ::= \blame{\blameLabel} \mid  \error\\
       \expra,\exprb \; & ::= 
                            b \mid \val \mid \var \mid 
                            \app{\expra}{\exprb} \mid
                            \tuple{\expra}{\expra} \mid 
                            \fst{\expra} \mid
                            \snd{\expra} \mid \\
                     & 
                           \coerce{\expra}{\seCa}  \mid 
                            \mkrefP{\expra} \mid
                            \derefP{\expra} \mid
                            \setrefP{\expra}{\exprb} \mid\\
                     &  
                           \mkrefM{\expra} \mid
                           \derefM{\expra} \mid
                           \derefTM{\expra}{\AllTa} \mid
                           \setrefM{\expra}{\exprb} \mid
                           \setrefTM{\expra}{\exprb}{\AllTa} 
\end{align*}
Figure~\ref{fig:semantics} defines the dynamic semantics. The language
forms $\mkrefP{\expra}$, $\derefP{\expra}$, and
$\setrefP{\expra}{\exprb}$ are for allocating, dereferencing, and
updating a proxied pointer, respectively.
The language form $\mkrefM{\expra}$ is for allocating a monotonic
reference. The form $\derefM{\expra}$ is for dereferencing a monotonic
reference with a fully static type whereas $\derefTM{\expra}{\AllTa}$
dereferences a monotonic reference that is partially or fully dynamic.
Similarly, $\setrefM{\expra}{\exprb}$ is for updating a monotonic
reference with a fully static type and
$\setrefTM{\expra}{\exprb}{\AllTa}$ is for updating a partially or
fully-dynamic monotonic reference.
Regarding the definition of values, the value form
$\coerced{\uncoerced}{\refPC{\seCa}{\seCb}}$ represents a proxied
reference whereas an address $a$ is a regular reference.

The dynamic semantics is given by three reduction relations: cast
reductions, program reductions, and state reductions. This
organization streamlines the treatment of monotonic references. The
heap $\mu$ maps an address to a value (for proxied references) or to a
value and a type (for monotonic references).  We refer to this type as
the run-time type information (RTTI) of the value and write
$\mu(a)_{\mathrm{val}}$ to access the value and
$\mu(a)_{\mathrm{rtti}}$ to access the RTTI.

The cast reduction rules define the semantics of applying a cast to a
value.  Space efficiency is achieved with the reduction that takes a
coerced value $\coerced{\uncoerced}{\seFinalC}$ and a coercion $\seCa$
and compresses the two coercions to produce the coerced value
$\coerced{\uncoerced}{\composeC{\seFinalC}{\seCa}}$.  There need not
be any cast reduction rules for proxied references, as a cast applied
to a reference is a value, similar to the function case. In contrast,
when a coercion is applied to a monotonic reference, we cast the
underlying value on the heap.  This cast is only allowed to make the
RTTI more precise. Any attempt to cast to an inconsistent type
triggers a runtime error.  Also, casting the underlying value causes
the heap to become an ``evolving heap'' which propagates the cast via
subsequent state transitions.

Regarding the program reduction rules, we have a different set of
reductions for operations on proxied versus monotonic references.  For
dereferencing a proxied reference, there are two rules, one for raw
addresses and the other for a proxy. Thus, an implementation must
dispatch on the kind of reference. If it's an address, the value is
loaded from the heap. If it's a proxy, the underlying reference is
dereferenced and then the proxy's read coercion $c$ is applied.  The
story is similar for writing to a proxied reference.
For monotonic references, there are two dereference operators.  For
the fully static dereference, we simply load the value from the
heap. For the partially dynamic dereference, we load the value from
the heap and cast it from it's RTTI to the expected type $T$. The
story is similar for writing to a monotonic reference.

Regarding the state reduction rules, the first rule simply enables
transitions according to the cast or program reduction rules.  The
next three rules propagate casts within an evolving heap. The first of
them commits the result of casting a value in the heap whereas the
second throws away the result of a cast in the case when the RTTI has
changed (due to cycles). The last rule handles the case when a cast
fails.

\section{The Grift Compiler}
\label{sec:implementation}

The Grift compiler takes programs in an extended version of the
gradually typed lambda calculus and compiles them to C, using the
Clang compiler to generate x86 executables. The Clang compiler
provides low level optimizations.
The first step in the Grift compiler is to translate to an
intermediate language with explicit casts. This process is
standard~\citep{Siek:2006bh,Siek:2008aa,Herman:2010aa} except for
optimizations to avoid unnecessary casts in dynamically typed code,
which we describe in Section~\ref{sec:opt-dyn}.

The next step in the compiler is exposing the runtime functions that
implement casts. We describe the representation of values
in~\ref{sec:value-rep}. We describe the implementation of typed-based
casts in Section~\ref{sec:type-based-implementation}, coercions in
Section~\ref{sec:coercions-implementation}, and monotonic references
in Section~\ref{sec:monotonic-references-implementation}.  After
lowering casts, Grift performs closure conversion using a flat
representation~\citep{Cardelli:1983aa,Cardelli:1984aa,Appel:1992fk},
and translates all constructors and destructors to memory allocations,
reads, writes, and numeric manipulation.

%



\subsection{Value Representation}
\label{sec:value-rep}

Values are represented according to their type.  An \IntT{} value is a
64-bit integer.  A \BoolT{} value is also a 64-bit integer, using the
C encoding of 1 for true and 0 for false.  For values of function
type, the representation depends on whether Grift uses type-based
casts or coercions.  In the former case, a function value is a pointer
to a closure.  A closure consists of 1) a function pointer, 2) a
pointer to a function for casting the closure, and 3) the values of
the free variables. In the later case (for coercions), a function
value is a pointer to one of two different kinds of closures and the
lowest bit of the pointer indicates which kind. The first kind, for
regular functions, is the same as above. The second kind which we call
a \emph{proxy closure}, is for functions that have been cast. It
consists of a function pointer (to a ``wrapper'' function), a pointer
to the underlying closure, and a pointer to a coercion.

A value of proxied reference type is a pointer to the data or to a
proxy. The lowest bit of the pointer indicates which. The
representation of the proxy depends on whether Grift uses type-based
casts or coercions. In the former case, the proxy consists of a
reference, the source and target types, and a blame label.  In the
later case, the proxy consists of a reference and a pointer to a
coercion (that is, a reference coercion).
A value of monotonic reference type is an address.

A value of type \DynT{} is a 64-bit integer, with the 3 lowest bits
indicating the type of the injected value. For atomic types
(e.g. \IntT{} and \BoolT{}), the injected value is stored in the other
61 bits.  For non-atomic types, the other 61 bits are a pointer to a
pair of the injected value and its type.

In the following, the macros for manipulating values have all uppercase
names to distinguish them from C functions. The macro definitions are
listed in Appendix~\ref{sec:macros}.




\subsection{Implementation of Type-based Casts}
\label{sec:type-based-implementation}

Type-based casts require a runtime representation of types.
Grift allocates all types known at compile time at the start of the
program. Each type is a 64 bit value, the lower 3 bits categorize whether it is
an atomic, function, proxied reference or monotonic reference type. For atomic
types, the other 61 bits indicate which atomic type. For function and reference
types, the other 61 bits point to a larger structure.  The structure for a
function type stores the arity, return type, and types for the parameters. The
structure for reference types consists of the referred-to type.


\begin{figure}[tbp]
\lstsetc%
\begin{lstlisting}
obj cast(obj val, type src, type tgt, blame lbl) {
  if (src == tgt) { return val; } 
  else if (src == DYN_TYPE) {
    return cast(UNTAG(val), TYPE(val), tgt, lbl); }
  else if (tgt == DYN_TYPE) { 
    return INJECT(val, src); }
  else if (TAG(src)==FUN_TAG && TAG(tgt)==FUN_TAG
           && ARITY(src)==ARITY(tgt))
     return UNTAG_FUN(val)->caster(val,src,tgt,lbl);
  else if (TAG(src)==REF_TAG && TAG(tgt)==REF_TAG) {
    type s = REF_TYPE(src), t = REF_TYPE(tgt);
    return MK_REF_PROXY(val, s, t, lbl);
  } else { raise_blame(lbl); }
}
\end{lstlisting}
\caption{The \castcall{} function.}
\label{fig:interp-cast}
\end{figure}

\paragraph{Casting Values}

Grift implements type-based casts with the C function named
\castcall{} (Figure~\ref{fig:interp-cast}) that takes a
value, two types (source and target), and a blame label, and returns a
value or signals an error. 
If the cast is between two identical types, then \castcall{}
returns the value unaltered.
If the source type is \DynT{}, then the underlying value and the type
are extracted and used to recursively cast to the target type of the
original cast.
Conversely, if the target type is \DynT{}, then the value is injected
into the representation for \DynT{}.


In case the source and target of the cast are function types,
\castcall{} wraps the value in a function that casts the arguments and
the return value appropriately.
For example, the cast:
\lstsetgrift
\centerline{\lstinline{(cast add1 (Int => Int) (Dyn => Dyn) L)}}
produces the wrapper function:

\centerline{\lstinline{(lambda (x) (cast (add1 (cast x Dyn Int L)) Int Dyn L))}}
\noindent
The wrapper function needs to have the same arity as the wrapped
function, so we find it convenient to attach to every closure, in
field \texttt{caster}, a function that knows how to generate the
appropriate wrapper function. This function allocates a closure whose
code pointer is to a wrapper function of the appropriate arity and
whose free variables provide access to the original closure and the
source and target types for the casts.
If the source and target type are inconsistent, an error is signaled
with a message that includes the blame label.

In case the source and target of the cast are proxied reference types,
\castcall{} allocates a reference proxy that stores the referred-to
types of the source and target and the underlying reference (which
could be either a pointer or another proxy).

To better understand the cost of a cast, Figure~\ref{fig:interp-cast}
colors the used macros and functions to indicate how expensive they are.
\lstsetc%
\lstinline{TAG} and \lstinline{UNTAG_FUN} perform pointer
and bit manipulation;
\lstinline{UNTAG}, \lstinline{TYPE}, and \lstinline{REF_TYPE} have a
few branches and perform loads from memory;
\lstinline{INJECT}, \lstinline{caster}, \lstinline{MK_REF_PROXY}
allocate memory;
and \lstinline{cast} is recursive but bound by the depth of types. 


\paragraph{Applying Functions}

For type-based casts, Grift doesn't need to distinguish between the
closures created by casting functions and closures created by defining
functions. As a result, the calling convention is simple and direct.
The generated code at each call site accesses the function pointer at
the beginning of the closure and performs an indirect call, passing
the closure to the function as an additional argument.

\begin{figure}[tb]
\lstsetc%
\begin{lstlisting}
obj ref_read(obj ref) { 
  if (TAG(ref) == REF_PROXY_TAG) {
    ref_proxy p = UNTAG_PREF(ref);
    obj v = ref_read(p->ref);
    return cast(v, p->src, p->tgt, p->lbl);
  } else { return *UNTAG_REF(ref); }
}
obj ref_write(obj ref, obj v) {
  if (TAG(ref) == REF_PROXY_TAG) {
    ref_proxy p = UNTAG_PREF(ref);
    obj cv = cast(v, p->tgt, p->src, p->lbl);
    return ref_write(p->ref, cv);
  } else { *UNTAG_REF(ref) = v; return UNIT; }
}
\end{lstlisting}
\caption{Code for reading and writing to references.}
\label{fig:guarded-write-read-lowered-type-based}
\end{figure}

\paragraph{Reading and Writing to Proxied References}

The C functions for reading and writing references are listed in
Figure~\ref{fig:guarded-write-read-lowered-type-based}.  The code for
reading dispatches on whether the reference is proxied or not, and if
proxied, recurses on the underlying reference (because it could be
another proxy).  When the recursion returns, the code casts the read
value from the source to the target type of the proxy. Otherwise the
read is accomplished by clearing the tag and dereferencing the
pointer. The story for writing to a reference is similar.

\subsection{Implementation of Coercions}
\label{sec:coercions-implementation}

The coercion implementation closely mirrors the semantics of coercions
given in Section~\ref{sec:semantics}. Coercions are represented as
heap allocated objects and the ones that are known at compile time are
allocated once at the start of the program. Grift calls the C function
\coercecall{}, described below, that interprets the coercion in
accordance with the semantics.


\paragraph{Coercion Representation.}

Similar to types, coercions are represented as 64-bit values where the
lowest 3 bits indicate whether the coercion is a projection,
injection, sequence, failure, or identity. The number of pointer tags
is limited, so the rest of the coercions are identified by a tag
stored in the first word of their structures. For an identity
coercion, the remaining 61 bits are not used. For the other coercions,
the remaining 61 bits stores a pointer to heap-allocated structure
that depends on the kind of coercion:

\begin{itemize}
\item Projection coercion (\prjC{T}{\blameLabel}) is represented in
  $2 \times 64$ bits: the first word is a pointer to the type $T$ of
  the projection and the second is a pointer to the blame label $p$.
\item Injection coercion (\injC{T}) is represented in 64 bits, holding
  a pointer to the type $T$ of the injected value.
\item Function coercion (\funC{c_1,\ldots,c_n}{c_r}) with $n$
  parameters is represented in $64(n + 2)$ bits, where the first word
  stores the secondary tag and arity, the second store a coercion on
  the return, and the remaining words store $n$ coercions for the
  arguments.
\item Proxied reference coercion (\refPC{c_1}{c_2}) is represented in
  $3\times 64$ bits, including the secondary tag, a coercion $c_1$ for
  writing, and another coercion $c_2$ for reading.
\item Monotonic reference coercion (\refMC{T}) is represented in
  $2\times 64$ bits, including the secondary tag and a type.
\item Sequences (\seqC{c_1}{c_2}) store two coercions in $2\times 64$
  bits.
\item Failure coercion (\failC{}{\blameLabel}{}) is represented in 64
  bits to store a pointer to the blame label.
\end{itemize}

\paragraph{Applying a Coercion}

\begin{figure}[tbp]
\lstsetc%
\begin{lstlisting}
obj coerce(obj v, crcn c) {
  switch(TAG(c)) {
  case ID_TAG: return v;
  case SEQUENCE_TAG:
    sequence seq = UNTAG_SEQ(c);
    return coerce(coerce(v, seq->fst), seq->snd);
  case PROJECT_TAG:
    projection proj = UNTAG_PRJ(c);
    crcn c2 = mk_crcn(TYPE(v), proj->type, proj->lbl);
    return coerce(UNTAG(v), c2);
  case INJECT_TAG:
    injection inj = UNTAG_INJECT(c); 
    return INJECT(v, inj->type); }
  case HAS_2ND_TAG: {
    sec_tag tag = UNTAG_2ND(c)->second_tag;
    if (tag == REF_COERCION_TAG) {
      if (TAG(v) != REF_PROXY) { 
        return MK_REF_PROXY(v, c); }
      else { ref_proxy p = UNTAG_PREF(v); 
        crcn c2 = compose(p->coercion, c);
        return MK_REF_PROXY(p->ref, c2); } 
    } else if (tag == FUN_COERCION_TAG)
      if (TAG(v) != FUN_PROXY) { 
        return UNTAG_FUN(v).caster(v, c); }
      else { fun_proxy p = UNTAG_FUN_PROXY(v); 
        crcn c2 = compose(p->coercion, c);
        return MK_FUN_PROXY(p->wrap, p->clos, c2); } 
  }
  case FAIL_TAG: raise_blame(UNTAG_FAIL(c)->lbl);
  }
}
\end{lstlisting}
\caption{The \coercecall{} function applies a coercion to a value.}
\label{fig:coercion-interp}
\end{figure}

The application of a coercion to a value is implemented by a C
function named \coercecall{}, shown in
Figure~\ref{fig:coercion-interp}, that takes a value and a coercion
and either returns a value or signals an error.
The \coercecall{} functions dispatches on the coercion's tag.
Identity coercions return the value unchanged.
Sequence coercions apply the first coercion and then the second
coercion.
Injection coercions build a value of type \DynT{}.
Projection coercions take a value of type \DynT{} and build a new
coercion from the runtime type to the target of the projection, which
it applies to the underlying value.

When coercing a function, \coercecall{} checks whether the
function has previously been coerced. If so, Grift builds a new proxy
closure by copying over the wrapper function and the underlying
closure, but its coercion is the result of composing the proxy's
coercion with the coercion being applied via
\composecall{} (Appendix \ref{sec:macros} Figure~\ref{fig:compose-coercion}).
If the function has not been previously coerced, then we access its
function pointer for casting and apply that to the function and the
coercion that needs to be applied. This ``caster'' function allocates
and initializes a proxy closure.
Coercing a proxied reference builds a proxy that stores two coercions,
one for reading and one for writing, and the pointer to the underlying
reference. In case the reference has already been coerced, the old and
new coercions are composed via \composecall{}, so that there will only
ever be one proxy on a proxied reference, which ensures space
efficiency.
Failure coercions halt execution and report an error using the blame
label.

\paragraph{Applying Functions}

Because the coercions implementation distinguishes between regular
closures and proxy closures, one might expect closure call sites to
branch on the type of closure being applied. However, this is not the
case because Grift ensures that the memory layout of a proxy closure is
compatible with the regular closure calling convention. The only change
from the type-based implementation's calling convention (described in
Section~\ref{sec:type-based-implementation}) is that we have to clear
the lowest bit of the pointer to the closure which distinguishes proxy
closures from regular closures. This representation is inspired by a
technique used in~\citet{Siek:2012uq} which itself is inspired
by~\citet{Findler:2002eu}.


\paragraph{Reading and Writing to Proxied References}

For the coercion implementation, Grift generates code for proxied
reference reads and writes that is similar to the code in the
type-based cast implementation. However, there are two slight
differences: since coercions are space-efficient, we know that there
will be at most one proxy, and therefore there is no need to recurse
on the proxied reference and the \coercecall{} function is used
with the coercion contained in a proxy if present.

\subsection{Implementation of Monotonic References}
\label{sec:monotonic-references-implementation}

A monotonic heap cell has two parts; the first stores runtime type
information (RTTI), and the second stores the value. Grift generates
pointer dereference and write instructions for reading and writing a
fully statically-typed monotonic reference. Otherwise, the value being
read or written has to be cast. The details of the latter process and
that of casting an address is described below.

\paragraph{Type-based Casts}

To cast a monotonic reference, we cast the underlying value on the
heap. First, the RTTI is read from the first word pointed to by the
address. The address is returned if the RTTI equals the target type
of the cast. However the equality check can be expensive if implemented
naively because the structures of both types will be traversed. Instead,
we hashcons~\cite{Allen:1978:AL:542865} all types to reduce structural
equality to pointer equality. If the check fails, we call
\inline{tglb}, which computes the greatest lower bound of two types,
and then overwrite the RTTI with the result. Next we call
\castcall{} on the value to cast it from the old RTTI to the new
RTTI. After \castcall{} returns, we check if the current RTTI is
the same as the one we wrote earlier to the heap and write the new value
to the heap only if they are indeed equal. Otherwise, a value with a
more precise type has already been written to the heap so we discard the
current value and return the address.

Reading from a non-static reference proceeds as follows: the value is
read from the second word pointed to by the address, the RTTI is read
from the first word, then \castcall{} is called on the value, the
RTTI, and the statically recorded type. For writing, the RTTI will be
read first from the heap and then \castcall{} will be called on the
new value, the statically recorded type and the RTTI. Again, we check if
the RTTI has changed during the casting process, if yes, we drop the
casted value, otherwise, we write the new value to the heap.

\paragraph{Coercions}

The story for coercions is similar. The only difference is that the
generated code builds coercions out of the RTTI and the other input type
with a call to \mkcrcncall{} and calls \coercecall{} instead
of \castcall{}.

\paragraph{Coercions and casts together?}
\label{sec:lazy-coercions}

An astute reader would notice that the implementation details of
coercing is very similar to casting in many scenarios with the
difference being creating a coercion out of the source and target
types and calling \coercecall{} instead of \castcall{}. Reading and
writing partially typed monotonic references is an obvious example of
such scenario where a coercion is created only to be immediately
consumed. Grift is clever in optimizing such cases by deferring
coercion creation until it is actually needed to be stored or composed
and uses the types to simulate coercions in other cases.  We refer to
this as \emph{lazy coercion creation}.  Our experiments show that this
optimization results in performance gains on the quicksort and n-body
benchmarks and no performance differences on the others.


\subsection{Specializing Casts and Coercions}
\label{sec:spec-cast}



Typically, types and coercions are inspected at runtime when values get
casted. However, for many casts, some or all of the types/coercions
involved are known at compile time. Grift can recognize such casts and
partially evaluate them, generating more efficient code. Consider the
following expression in the intermediate language where casts are
explicit:
\lstsetgrift
\centerline{\lstinline{(cast n Dyn Int l)}}

\noindent
a straight forward compilation is a \castcall{} call which
dispatches on types at runtime. But with specialization, it will get
compiled to the following efficient code:

%
\lstsetgrift
\centerline{\lstinline{(TAG(n) == INT_TYPE) ? UNTAG_INT(n) : raise_blame(l)}}

\noindent which is basically the body of the branch inside
\castcall{} where the type arguments are \DynT{} and \IntT{}.

While lazy coercion creation reduces the number of coercions that get
allocated dynamically in certain cases, specialization simplifies the
coercions that are known at compile time.  Our experiments show that
specializing coercions significantly improves some benchmarks, such as
Matrix Multiplication, but only slightly improves the rest
(Appendix~\ref{sec:appendix-specialization}).

\subsection{Optimizing Dynamically Typed Code}
\label{sec:opt-dyn}

\lstsetgrift The straightforward approach to inserting
casts~\citep{Siek:2015ac} can cause unnecessary overhead in
dynamically typed regions of code. Consider the function
\lstinline{(lambda (f) (f 42))} which applies a function \lstinline{f}
injected into \DynT{}. The straightforward cast insertion would
compile it to:
\begin{lstlisting}
(lambda (f : Dyn)
  ((cast f Dyn (Dyn => Dyn) L) (cast 42 Int Dyn L)))
\end{lstlisting}
The cast on \lstinline{f} will allocate a function proxy if the source
type of \lstinline{f} is not \lstinline{(Dyn => Dyn)}. Although
allocating a proxy is important to maintain the semantics, the
allocation is unnecessary in this case because it will be consumed right
away. Instead, Grift specializes these cases by generating code that
does what a proxy would do without allocating one. Grift applies this
optimization to proxied references and tuples as well.

\section{Performance Evaluation}
\label{sec:performance-eval}

In this performance evaluation, we seek to answer a number of research
questions regarding the runtime overheads associated with gradual
typing.

\begin{enumerate}
\item \textbf{What is the overhead of gradual typing?}
  (Sec.~\ref{sec:external-comparison}) We subdivide this question into
  the overheads on programs that are (a) statically typed,
  (b) dynamically typed, and (c) partially typed.
  \label{RQ:overhead}
  \label{RQ:static}
  \label{RQ:dyn}
  \label{RQ:partial}
  


\item \textbf{What is the cost of achieving space efficiency with
  coercions?} (Section~\ref{sec:cost-space-efficiency})
  \label{RQ:cost-space}

\item \textbf{How do monotonic references compare with proxied references?}
  (Section~\ref{sec:monotonic-versus-proxied})
  \label{RQ:mono-vs-proxy}
\end{enumerate}


\subsection{Experimental Methodology}
\label{sec:methods}

In these experiments we use benchmarks from a number of sources: the
well-known Scheme benchmark suite (R6RS) used to evaluate the
Larceny~\citep{Hansen:2002aa} and Gambit~\citep{Feeley:2014aa}
compilers, the PARSEC benchmarks~\citep{Bienia:2008aa}, and the
Computer Language Benchmarks Game (CLBG).
We do not use all of the benchmarks from these suites due to the
limited number of language features currently supported by the Grift
compiler. We continue to add benchmarks as Grift grows to support more
language features.
In addition to the above benchmarks, we also include two textbook
algorithms: matrix multiplication and quicksort.
We chose quicksort in particular because it exhibits space efficiency
issues similar to the benchmarks
of~\citet{Takikawa:2015aa,Takikawa:2016aa}.
The benchmarks that we use are:
\begin{description}
\item[tak] (R6RS) This benchmark, originally a Gabriel benchmark, is a
  triply recursive integer function related to the Takeuchi
  function. It performs the call \texttt{(tak 40 20 12)}. A
  test of non-tail calls and arithmetic.
\item[ray] (R6RS) Ray tracing a simple scene, 20 iterations. A test of
  floating point arithmetic. Adapted from Example 9.8 of Paul Graham's
  book on ANSI Common Lisp.
\item[blackscholes] (PARSEC) This benchmark, originally an Intel RMS
  benchmark, calculates the prices for a portfolio of European options
  analytically with the Black-Scholes partial differential
  equation. There is no closed-form expression for the Black-Scholes
  equation and as such it must be computed numerically.
\item[matmult] (textbook) A triply-nested loop for matrix
  multiplication, with integer elements.  The matrix size is
  $400\times 400$ in the comparisons to other languages
  (Sec.~\ref{sec:eval-static} and \ref{sec:eval-dynamic}) and
  $200\times 200$ for the evaluation of partially typed programs
  (Sec.~\ref{sec:cost-space-efficiency} and
  \ref{sec:monotonic-versus-proxied}).
\item[quicksort] (textbook) The standard quicksort algorithm on
  already-sorted (worst-case) input, with integer arrays of size
  $10,000$ in the comparison to other languages and $1,000$ for
  the partially typed programs.
\item[fft] (R6RS) Fast Fourier Transform on $65,536$ real-valued
  points. A test of floating point.
\item[n-body] (CLBG) Models the orbits of Jovian planets, using a
  simple symplectic-integrator.
\end{description}

\paragraph{Experimental Setup}

All experiments were conducted on an unloaded machine with a 4-core
Intel(R) Core(TM) i7-4790 CPU @ 3.60GHz processor with 8192 MB of
cache and 16 GB of RAM running Red Hat 4.8.5-16.  The C compiler we
used is Clang 5.0.0, the Gambit compiler is version 4.8.8, Racket is
version 6.10.1, and Chez Scheme is version 9.5.1. All time
measurements use real time, and 30 samples were collected of each time
measurement and the mean reported.

\paragraph{Measuring the Performance Lattice}
\label{sec:perf-lattice}

\citet{Takikawa:2016aa} observe that evaluating the performance of
implementations of gradually typed languages is challenging because
one needs to consider not just one version of each program, but the
very many versions of a program that can be obtained by
adding/removing static types.  For languages with \emph{coarse-grained
  gradual typing}, as in \citet{Takikawa:2016aa}, one considers all
the combinations of making each module typed or untyped, so there are
$2^m$ configurations of the $m$ modules. The situation for languages
with \emph{fine-grained gradual typing}, as is the case for Grift, is
considerably more difficult because any type annotation, and even any
node within a type annotation, may be changed to \DynC{}, so there are
millions of ways to add type annotations to these benchmarks.

For our experiments on partially typed programs, we randomly sample 90
configurations from across the spectrum of type annotations for each
benchmark.  Our sampling approach starts from a statically-typed
program, and at each type annotation in the program, we generate the
array of all types that are less precise than that type. We then
sample a program from the lattice at a certain percentage of type
annotation by generating random indices into these arrays, use them to
choose less precise types, and finally, insert the chosen types into
the locations of the original type annotations in the benchmark. The
random indices are generated in a controlled way to make sure the
percentage of type annotations in the resultant sample will fall
within the range we are aiming for.

This algorithm ensures that when sampling to produce programs with a
particular percentage of type annotations, all programs that satisfy
the type percentage constraint are equally likely to be
chosen. However, the algorithm suffers from large memory
consumption. At this time we are unable to run it on the ray tracing
benchmark because of the size of its types. As such, we have omitted
ray from the partially typed comparisons.

\subsection{Gradual Typing Overhead and Comparison}
\label{sec:external-comparison}

The purpose of this section is to answer research question
\ref{RQ:overhead}, i.e., what is the overhead of gradual typing?  Of
course, to ultimately answer this question one would need to consider
all possible implementations. So the actual question we answer is:
what is the overhead of gradual typing in an ahead-of-time compiler
that implements space-efficient coercions and monotonic references?
To answer this question, it is important to isolate the overheads of
gradual typing from other sources of overhead. Thus, we have
implemented a variant of the Grift compiler, named Static Grift, that
requires the input program to be statically typed and does not
generate any code in support of gradual typing. We then compare
(gradually typed) Grift with Static Grift.

Of course, one threat to validity is the possibility that the
performance of Static Grift could be so poor that the overheads due to
gradual typing are drowned out. To allay this fear, we include
external comparisons to other programming language implementations,
including statically typed languages such as OCaml, and dynamically
typed languages such as Scheme and Racket. The upshot of this
comparison is that the performance of Grift is in the same ballpark as
these other implementations. It is tempting to interpret these
comparisons in a competitive light, but we encourage the reader to
abstain from such thoughts. The point is not that we would convince
the reader to start compiling programs with Grift (you shouldn't
because it does not support a full language), but to convince
implementors of gradually-typed languages that coercions and monotonic
references are worthwhile.

Finally, we note that the semantics of monotonic references is
different from proxied references, so the Grift compiler implements
two different gradually typed languages. We refer to them as Proxied
Grift and Monotonic Grift. Both versions use coercions (not type-based
casts), specialize casts, and perform lazy coercion creation, as
described in Section~\ref{sec:implementation}.  In this section we
report on the performance of both Proxied Grift and Monotonic Grift.

\subsubsection{Evaluation on Statically Typed Programs}
\label{sec:eval-static}

\begin{figure}[tbp]
  \centering
  \includegraphics[scale=.20]{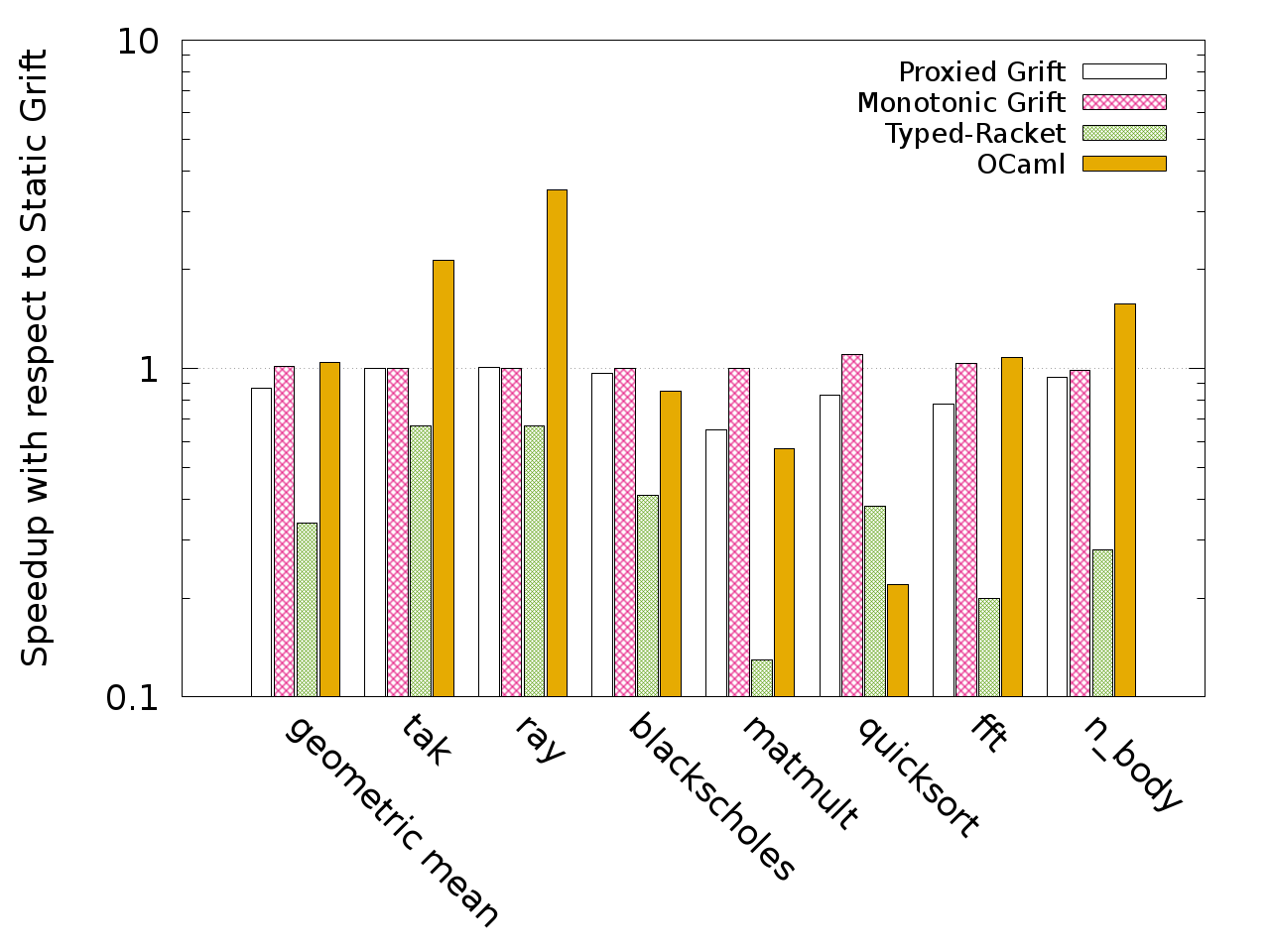}
  \vspace{-10pt}
  \caption{A comparison of the speedup on statically typed programs of
    Proxied Grift and Monotonic Grift with respect to Static Grift.
    For context, we also include the speedup of Typed Racket and
    OCaml.
    }
  \label{fig:fully-static-suite-plot}
\end{figure}

Figure~\ref{fig:fully-static-suite-plot} shows the results of
evaluating the speedup of Proxied and Monotonic Grift with respect to
Static Grift on statically typed versions of the benchmarks.  We see
that the performance of Monotonic Grift is no lower than $0.99\times$
of Static Grift on all the benchmarks whereas the performance of
Proxied Grift sometimes dips to $0.65\times$ that of Static Grift.
To put the performance of Grift in context, it is comparable to OCaml
and better than fully static Typed Racket.
\begin{quote}
\textbf{Answer to research question (\ref{RQ:static} a):} the
overhead of gradual typing for statically typed code is consistently
low with monotonic references but sometimes high with proxied
references.
\end{quote}





To mitigate the differences between runtime initialization we use
internal timing. For Type Racket we make sure to use the floating
point specialized math operations, but since there is no safe and
well-performing equivalent operation for fixed width integers we are
forced to use the polymorphic math operators for integers. We also
make no attempt to account for differences in garbage collection.



\subsubsection{Evaluation on Dynamically Typed Programs}
\label{sec:eval-dynamic}

\begin{figure}[tbp]
  \centering
  \includegraphics[scale=.20]{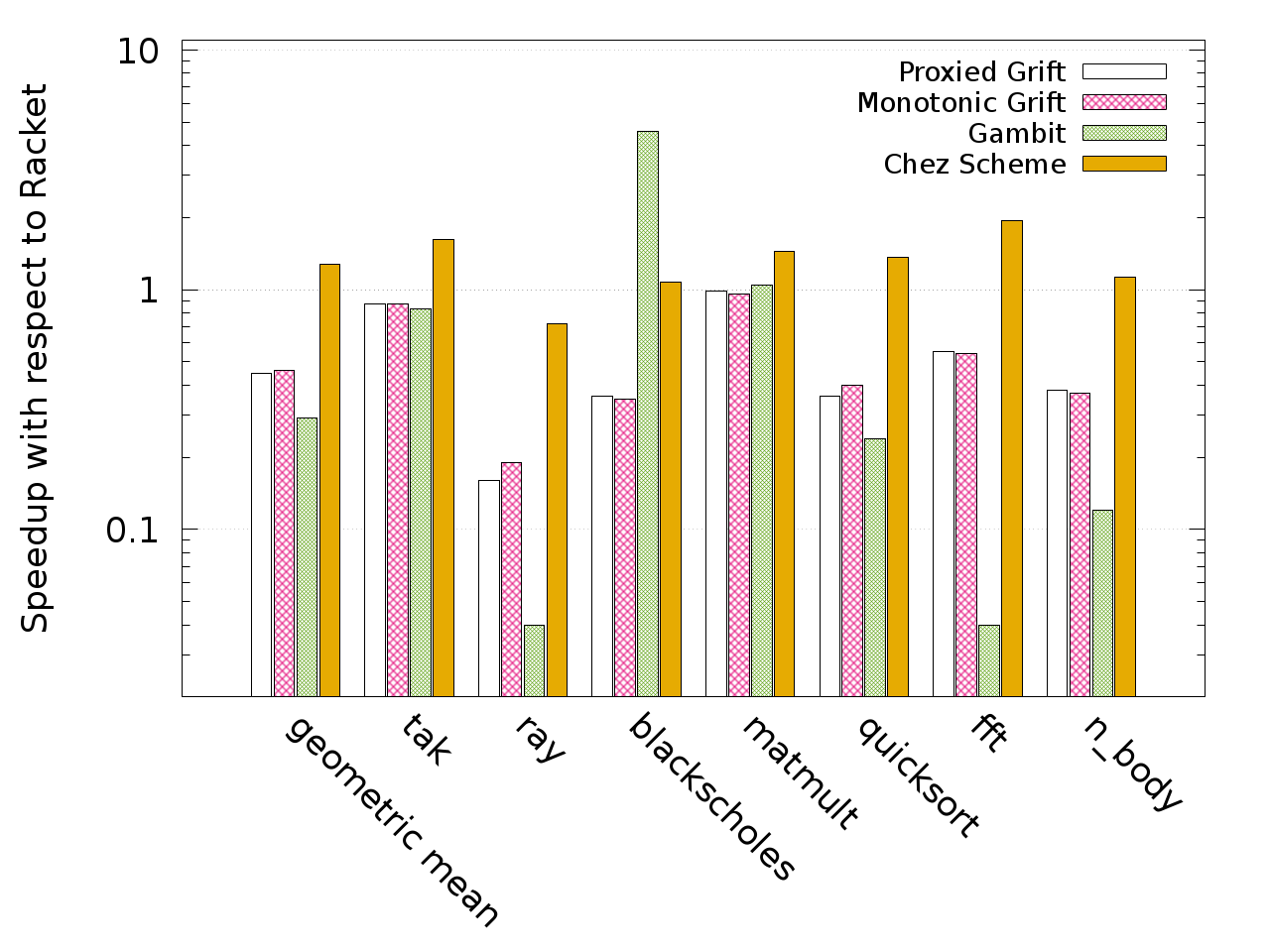}
  \vspace{-10pt}
  \caption{A comparison of the speedup on dynamically typed programs
    of Proxied Grift and Monotonic Grift with respect to Racket.
    For context, we also include the speedup of 
    other Scheme implementations (Gambit and Chez Scheme).
  }
  \label{fig:fully-dyn-suite-plot}
\end{figure}

Figure~\ref{fig:fully-dyn-suite-plot} shows the results of evaluating
the speedup of Proxied and Monotonic Grift with respect to Racket on
dynamically typed versions of the benchmarks.  Note that we have not
implemented a ``Dynamic Grift'' (analogous to Static Grift, but for
dynamic languages), but instead we compare to Racket. 
The figure also includes results for Gambit and Chez Scheme.  We see
that the performance of Grift is generally higher than Gambit but
lower than Racket and Chez Scheme on these benchmarks.
This experiment does not tease apart which of these performance
differences are due to gradual typing per se and which of them are due
to orthogonal differences in implementation, e.g., ahead-of-time
versus JIT compilation, quality of general-purpose optimizations, etc.
Thus we draw the following conservative conclusion.
\begin{quote}
\textbf{Answer to research question (\ref{RQ:dyn} b)} the overhead of
gradual typing for dynamically typed code is currently reasonable but
there is still some improvement to be made.
\end{quote}

\subsubsection{Evaluation on Partially Typed Programs}

\begin{figure}[tbp]
  \centering
  \begin{subfigure}[b]{0.5\textwidth}
    \centering
    \includegraphics[scale=.19]{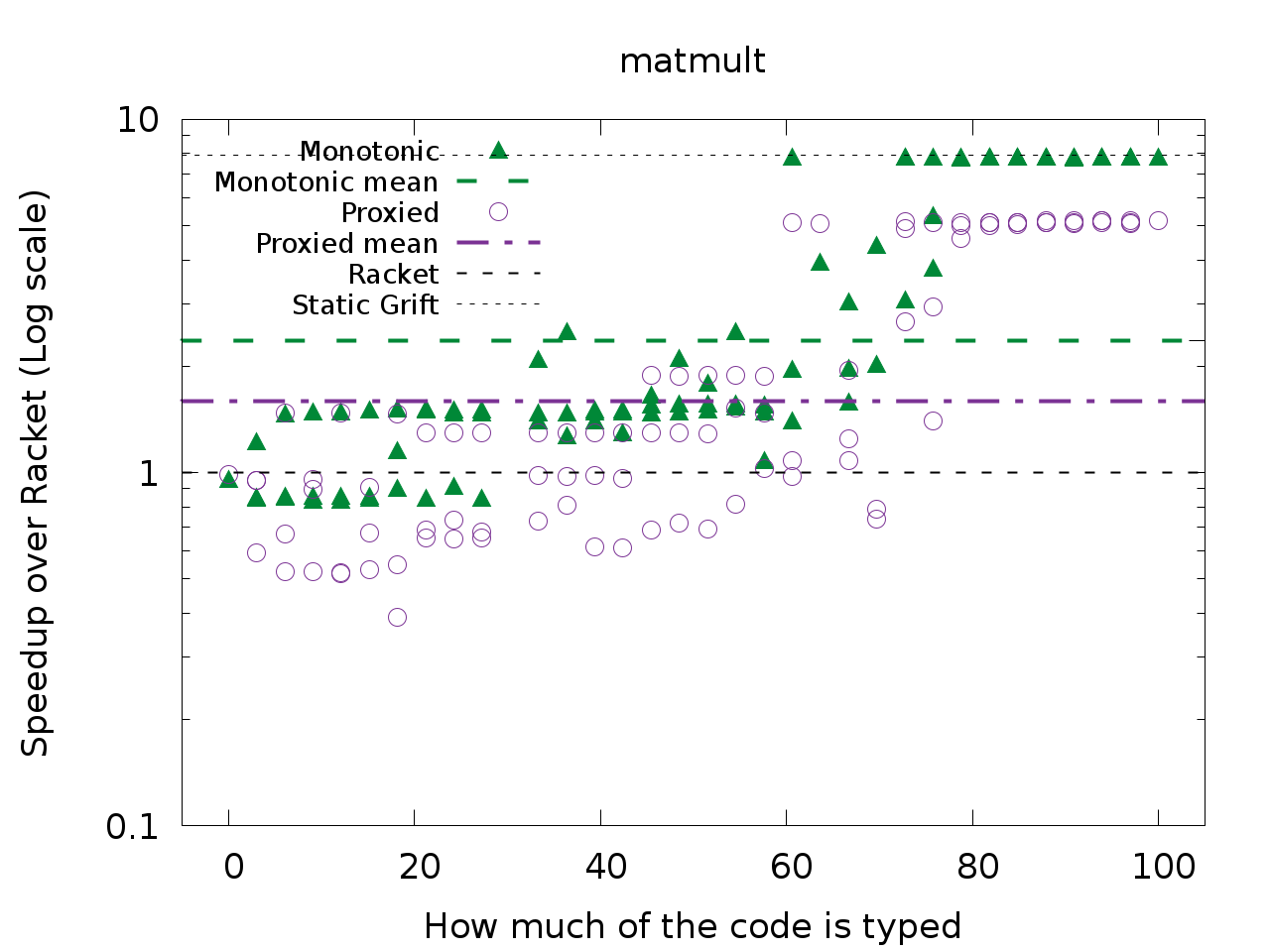}
    \label{fig:perflattice-matmult-casts-plot}
  \end{subfigure}
  \begin{subfigure}[b]{0.5\textwidth}
    \centering
    \includegraphics[scale=.19]{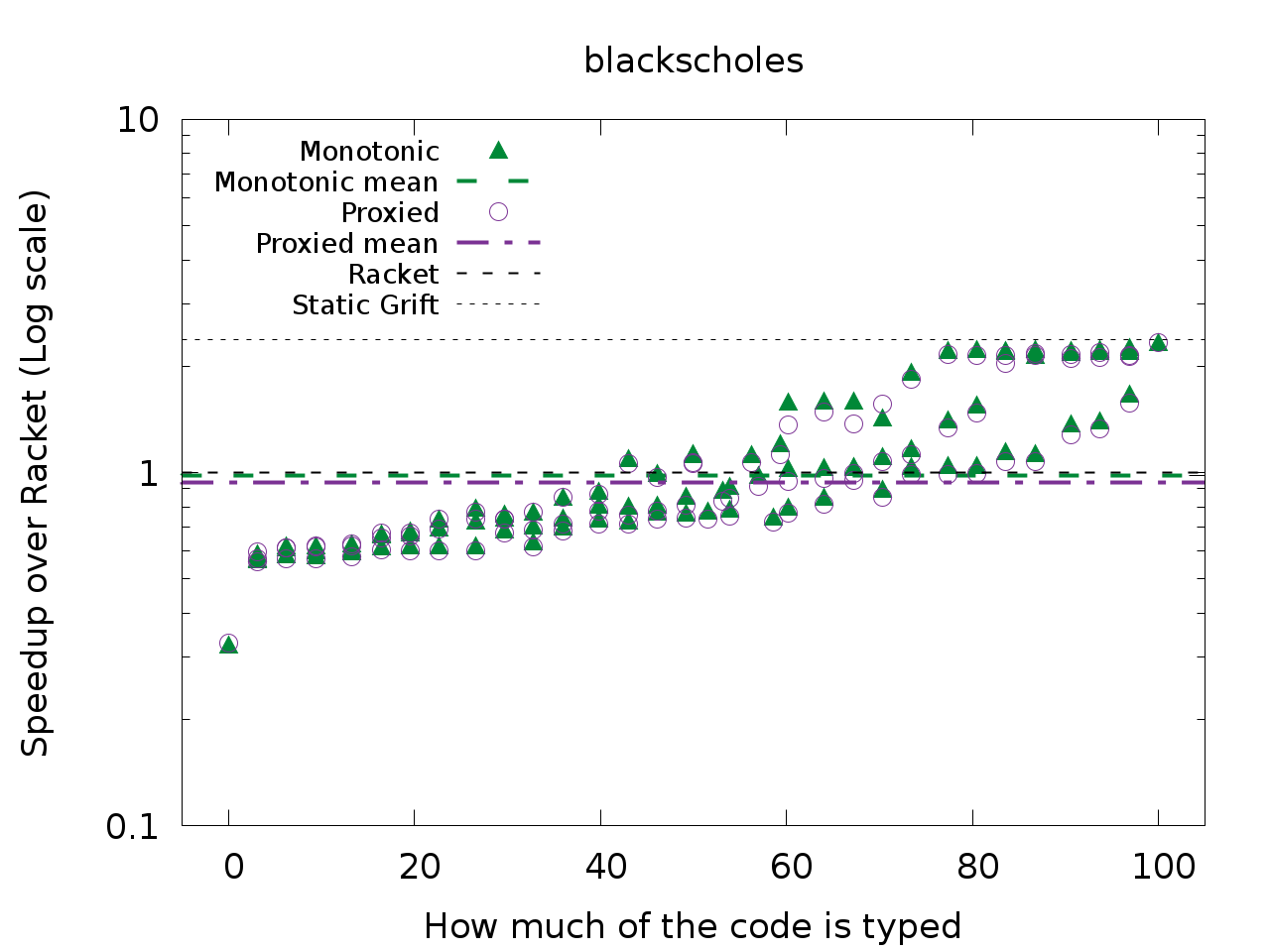}
    \label{fig:perflattice-blackscholes-worstcase-casts-plot}
  \end{subfigure}
  \begin{subfigure}[b]{0.5\textwidth}
    \centering
    \includegraphics[scale=.19]{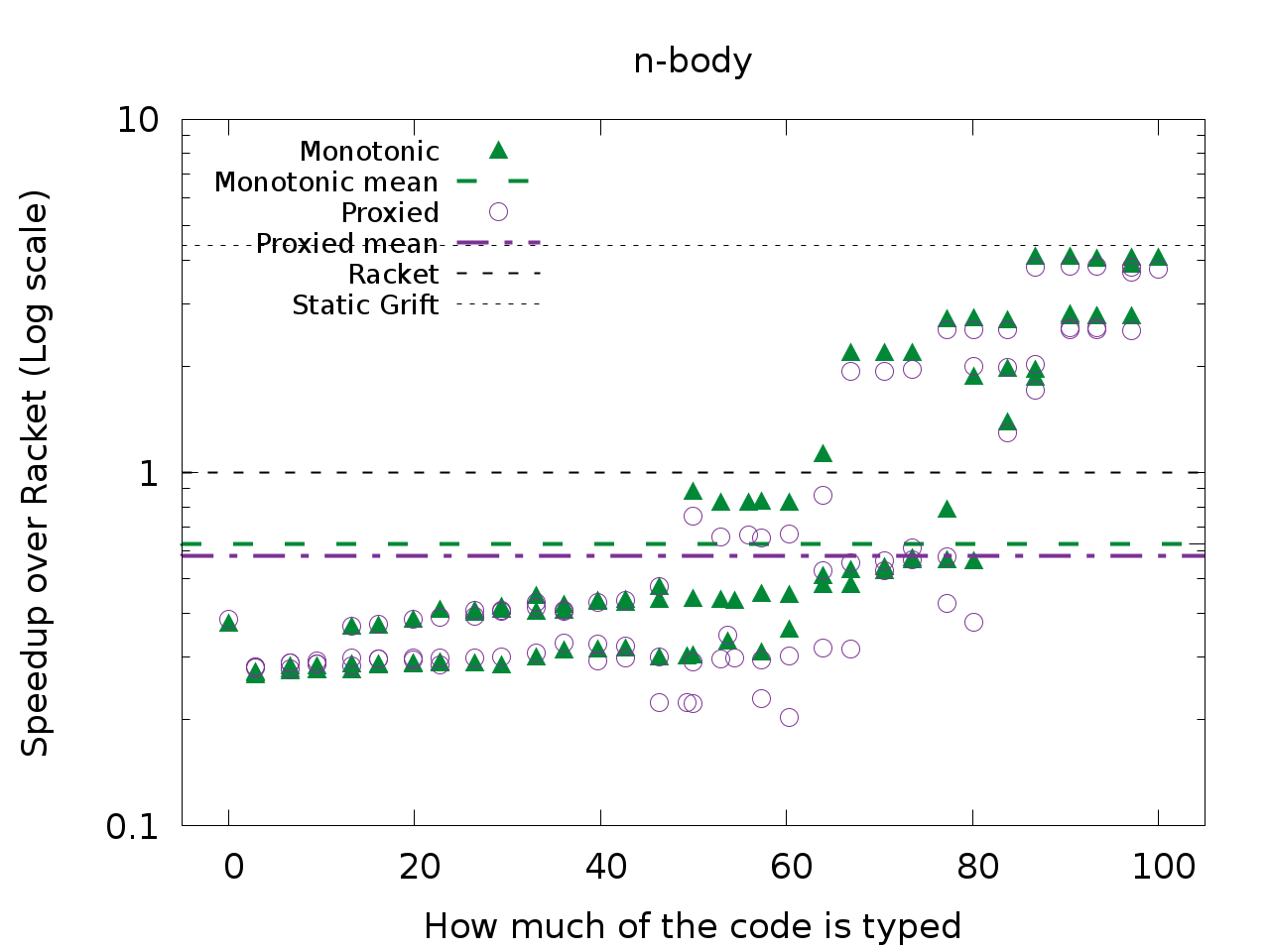}
    \label{fig:perflattice-nbody-casts-plot}
  \end{subfigure}
  \caption{Evaluation of Monotonic Grift and Proxied Grift on
    partially typed programs.  The y-axis is speedup relative to
    Racket on a logarithmic scale. The x-axis varies the amount of
    type annotations in the program, from 0\% to 100\%.  On the whole,
    monotonic references are more efficient than proxied references in
    partially typed code.
  }
\label{fig:performance-lattices-casts}
\end{figure}

Figure~\ref{fig:performance-lattices-casts} shows the speedup of
Proxied and Monotonic Grift with respect to Racket on a large number
of partially typed configurations of each benchmark (recall that
Section~\ref{sec:perf-lattice} discusses the selection of
configurations). The x-axis varies the amount of type annotations in
the program, from 0\% on the left to 100\% on the right.  The
performance of Racket is the horizontal line at $1$. We also include a
horizontal line for Static Grift, which represents the best
performance one could hope for. In
Figure~\ref{fig:performance-lattices-casts} we show the results for
three benchmarks.
We select matmult,
blackscholes, and n-body as representatives of best, middle, and worst
case comparisons to Racket.

In matmult, the performance of Monotonic Grift is slightly below
Racket for the untyped configurations but then climbs to nearly
$10\times$ speedup for configurations that are $80\%$ typed or more.
The performance of Proxied Grift trails that of Monotonic, but the
trend is similar. Note that the mean for Proxied Grift (purple
horizontal line) and for Monotonic Grift (green horizontal line) are
both well above Racket.

In blackscholes, Grift starts around $0.3\times$ the speed of Racket
and then gradually climbs to match Racket at $60\%$ typed and then
exceeds Racket by about $2\times$ at $80\%$ typed. The mean for Grift
is similar to Racket on this benchmark.

In n-body, Grift again starts out slower than Racket but becomes
faster once the configurations are $80\%$ typed.  However, in this
benchmark the mean for Grift is significantly lower, around
$0.6\times$ the speed of Racket.
\begin{quote}
\textbf{Answer to research question (\ref{RQ:dyn} c)} the overhead of
gradual typing on partially typed code is currently reasonable but
there is still some improvement to be made.
\end{quote}

\subsection{The Runtime Cost of Space Efficiency}
\label{sec:cost-space-efficiency}

\begin{figure}[tbp]
  \centering
  \begin{subfigure}[b]{0.5\textwidth}
    \centering
    \includegraphics[scale=.19]{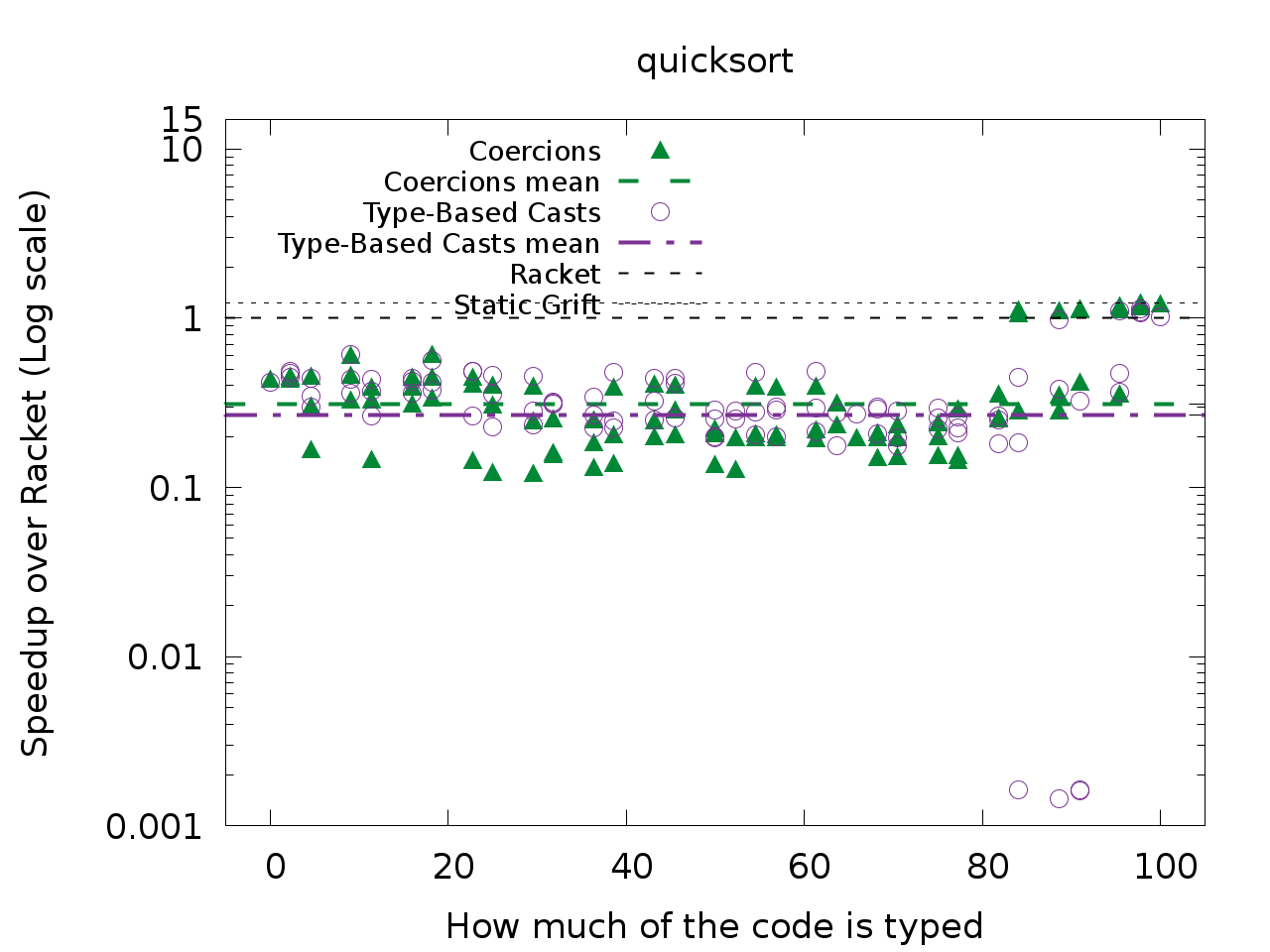}
    \label{fig:perflattice-quicksort-worstcase-guarded-plot}
  \end{subfigure}
  \begin{subfigure}[b]{0.5\textwidth}
    \centering
    \includegraphics[scale=.19]{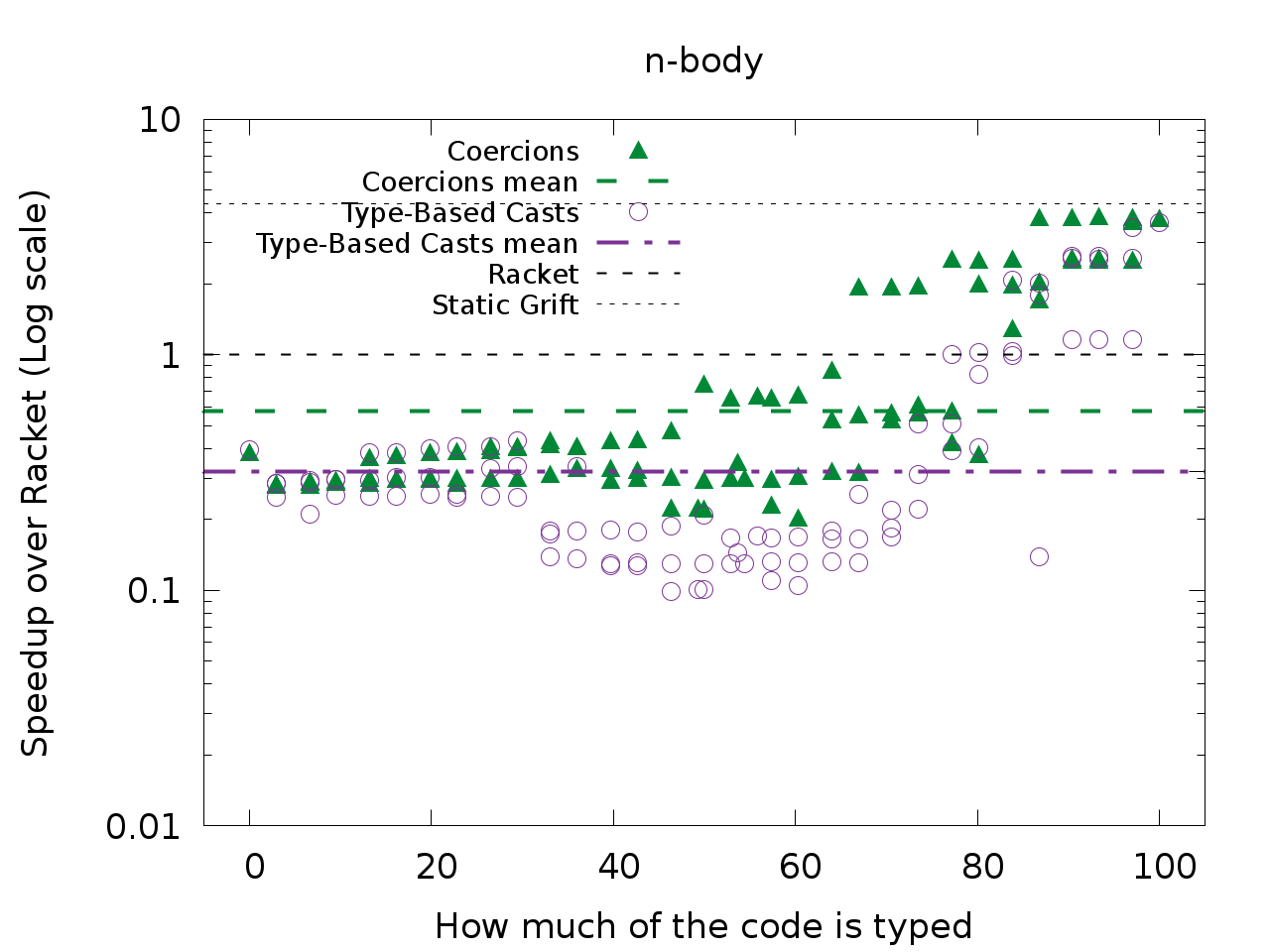}
    \label{fig:perflattice-nbody-guarded-plot}
  \end{subfigure}
  \begin{subfigure}[b]{0.5\textwidth}
    \centering
    \includegraphics[scale=.19]{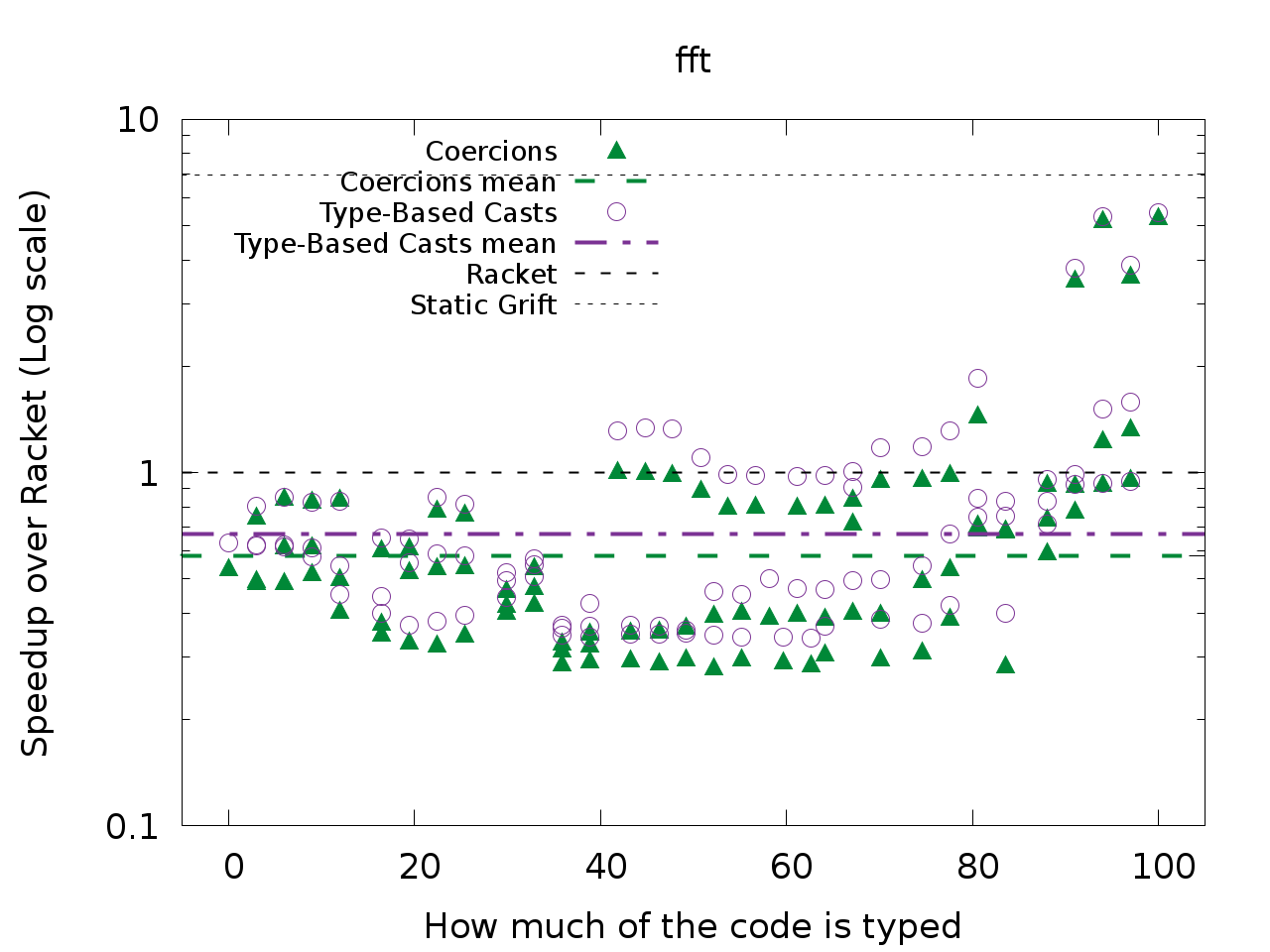}
    \label{fig:perflattice-fft-guarded-plot}
  \end{subfigure}
  \caption{ We compare Grift with coercions to Grift with type-based
    casts (both use proxied references) across partially
    typed-programs to evaluate the cost of space-efficiency. The
    y-axis is speedup over Racket on a log scale. The x-axis is the
    percentage of types annotations, from 0\% to 100\%.  Note that some
    configurations of quicksort exhibit catastrophic slowdowns with
    type-based casts.
    }
\label{fig:performance-lattices-guarded}
\end{figure}

In Figure~\ref{fig:performance-lattices-guarded} we compare the
performance of Grift with type-based casts, to Grift with coercions,
which are space-efficient. We compare these two approaches on partially
typed configurations of the benchmarks.
The quicksort benchmark is the only one that elicits space efficiency
problems. In that benchmark, type-based casts exhibit catastrophic
performance on some configuration (the circles at the bottom, between
$0.001$ and $0.002$). In fact, in some configurations, the time
complexity of quicksort changes from $O(n^2)$ to $O(n^3)$!  This
occurs when the array is cast on each recursive call, and each cast
adds another proxy that induces overhead on subsequent reads and
writes to the array.

On the other hand, the coercion-based approach successfully eliminates
these catastrophic slowdowns.  In benchmarks that do not elicit space
efficiency problems, we see a general trend of the coercions being
roughly equal to the performance type-based casts.  Across all
benchmarks the speedup of coercions over type-based cast is between
$0.82\times$ (the fft benchmark) and $1.93\times$ (the matmult
benchmark).

\begin{quote}
\textbf{Answer to research question \eqref{RQ:cost-space}:}
Space-efficient coercions offer a ``pay as you go'' cost model.  On
benchmarks without space-efficiency issues, we sometimes see a mild
speedup and sometimes a mild slowdown.  However, where space
efficiency is needed, coercions eliminate the catastrophic slowdowns.
\end{quote}



\subsection{Monotonic versus Proxied References}
\label{sec:monotonic-versus-proxied}


We return our attention to
Figure~\ref{fig:performance-lattices-casts}, but this time with an eye
towards evaluating whether monotonic references perform better than
proxied references. Indeed, monotonic references are faster by
$\MonotonicVsProxiedMatmult\times$ in matmult and by
$\MonotonicVsProxiedNbody\times$ in n-body. Additionally, the
experiment shows that monotonic references match the performance of
Static Grift in cases where the benchmark source code is closer to be
fully typed.

\begin{quote}
\textbf{Answer to research question \eqref{RQ:mono-vs-proxy}:}
Monotonic references are more efficient than proxied ones on partially
typed programs and enable especially low overhead in statically typed
code.
\end{quote}

\section{Conclusion}



We have presented Grift, a compiler for exploring implementations of
gradually typed languages.
In particular, we implement several technologies that enable gradual
typing: type-base cast, space-efficient coercions, traditional proxied
references, and monotonic references.
Our experiments show that Grift with monotonic references is
competitive with OCaml on statically typed code.
For dynamically typed code, Grift is on par with Scheme
implementations using both proxied and monotonic implementations.
On partially typed code, our experiments show that coercions eliminate
the catastrophic slowdowns caused by space-inefficient casts.
Furthermore, we see significant speedups ($10\times$) as
60\% or more of a program is annotated with types.
Future work remains to improve the efficiency of coercions.

\clearpage
\pagebreak

\bibliography{all,extra}

\clearpage
\pagebreak

\appendix

\section{Values, Macros, and Compose}
\label{sec:macros}
\lstsetc
Figure~\ref{fig:value-rep} lists the C structs use to represent
values. Figure~\ref{fig:macros} lists the macros for manipulating
values. Figure~\ref{fig:compose-coercion} show the code for
\composecall{} runtime function which directly follows the
equations for compose in Figure~\ref{fig:semantics}.

\begin{figure*}[h]
\begin{lstlisting}
/* Types */
typedef type* ref_type;
typedef struct {int64_t arity; type ret; type args[]} fun_type;
typedef union {int64_t atm; ref_type* ref; fun_type* fun;} type;
/* Coercions */
typedef struct {type to; blame info;} project_crcn;
typedef type* inject_crcn;
typedef struct {crcn fst; crcn snd;} seq_crcn;
typedef struct {snd_tag second_tag; int32_t arity; crcn ret; crcn args[]} fun_crcn;
typedef struct {snd_tag second_tag; crcn write; crcn read} pref_crcn;
typedef struct {snd_tag second_tag; crcn rtti} mref_crcn;
typedef struct {char* lbl} fail_crcn;
#define ID NULL
/* Values */
#ifdef TYPE_BASED_CASTS
  typedef struct {obj* ref; type source; type target blame info;} ref_proxy;
  typedef struct {void* code;  (obj)(*caster)(obj, type, type, blame);  obj[]; } closure;
#elseif COERCIONS
  typedef struct {obj* ref; crcn cast;} ref_proxy
  typedef struct {void* code; (obj)(*caster)(obj, type, type, blame);
                    union {crcn  coerce; obj[] fvars;} } closure;
#endif
typedef struct {obj value; type source} nonatomic_dyn;
typedef union {int64_t atomic; nonatomic_dyn*} dynamic;
typedef union {int64_t fixnum; double  flonum; closure* clos; dynamic dyn} obj;
\end{lstlisting}
\caption{Value representations}
\label{fig:value-rep}
\end{figure*}

\begin{figure*}[h]
\begin{lstlisting}
/* All allocated values have 3 bits that can be used for tagging */
#define TAG(value) (((int64_t)value)&0b111)
#define UNTAG_INT(value) (((int64_t)value)&~0b111)
#define TAG_INT(value, tag)   (((int64_t)value)|tag)
/* Macros that manipulate types */
#define HEAD(type) (TAG(type))
#define ARITY(type) (((fun_type)UNTAG_INT(type))->arity)
#define REF_TYPE(type) (*((ref_type)UNTAG_INT(type)))
/* Macros that manipulate values in the obj union */
#define UNTAG_REF(ref) ((obj*)UNTAG_INT(ref))
#ifdef TYPE_BASED_CASTS
#define UNTAG_FUN(fun) ((closure*)(fun)))
#define MK_REF_PROXY(v, s, t, l) (tmp_rp = (ref_proxy*)GC_MALLOC(RP_SIZE), tmp_rp->value=v, \
      tmp_rp->source=s,tmp_rp->target=t,tmp_rp->info=l, (obj)TAG_INT(tmp_rp, REF_PROXY_TAG)
#elseif COERCIONS
#define UNTAG_FUN(fun) ((closure*)UNTAG_INT(fun))
#define MK_REF_PROXY(v, c) (tmp_rp = (ref_proxy*)GC_MALLOC(RP_SIZE), tmp_rp->value=v, \
      tmp_rp->coerce=c, (obj)TAG_INT(tmp_rp, REF_PROXY_TAG)
#endif 
/* Macros that manipulate values in the dynamic union */
#define UNTAG_NONATOMIC(value) ((nonatomic_dyn)UNTAG_INT(value))
#define UNTAG(v) ((TAG(v) == INT_TAG) ? (obj)(UNTAG_INT(v)>>3) : \
                     (TAG(v) == UNIT_TAG) ? (obj)UNIT_CONSTANT : ... (obj)UNTAG_NONATOMIC(v).value)
#define TYPE(v) ((TAG(v) == INT_TAG) ? (type)INT_TYPE : (TAG(v) == UNIT_TAG) ? (type)UNIT_TYPE : ... \
                    UNTAG_NONATOMIC(v)->source)
#define INJECT(v, s) ((s==INT_TYPE) ? TAG_INT(v<<3, INT_TAG) : (source==UNIT_TYPE) ? DYN_UNIT_CONSTANT : ... \
    (tmp_na = (nonatomic_dyn*)GC_MALLOC(NA_DYN_SIZE), tmp_na->value=value, tmp_na->source=s, (obj)tmp_na)
/* Macros that manipulate types in the crcn union */
#define UNTAG_2ND(c) ((struct {snd_tag second_tag;}*)UNTAG_INT(c))
/* UNTAG_PRJ, UNTAG_FAIL, UNTAG_SEQ are similar to UNTAG_INJ */
#define UNTAG_INJ(inj) ((inject_crcn)UNTAG_INT(inj))
/* MK_SEQ, MK_PROJECTION, MK_INJECTION are similar */
#define MK_REF_COERCION(r, w) (tmp_rc = (ref_crcn*)GC_MALLOC(RC_SIZE), tmp_rc->second_tag=REF_COERCION_TAG,\
    tmp_rc->read=r, tmp_rc->write=w, (crcn)(TAG_INT(tmp_rc, HAS_2ND_TAG)))    
\end{lstlisting}

\caption{Macros for manipulating values}
\label{fig:macros}
\end{figure*}

\begin{figure*}[h]
\lstsetc
\begin{lstlisting}
crcn compose(crcn fst, crcn snd) {
  if (fst == ID) { return snd; }
  else if (snd == ID) { return fst; } 
  else if (TAG(fst) == SEQUENCE_TAG) {
    sequence s1 = UNTAG_SEQ(fst);
    if (TAG(s1->fst) == PROJECT_TAG) { 
      return MK_SEQ(s1->fst, compose(s1->snd, snd)); }
    else if (TAG(snd) == FAIL_TAG) { return snd; }
    else { sequence s2 = UNTAG_SEQ(snd); 
      type src = UNTAG_INJ(s1->snd)->type; 
      type tgt = UNTAG_PRJ(s2->fst)->type; 
      blame lbl = UNTAG_PRJ(s2->fst)->lbl;
      crcn c = mk_crcn(src, tgt, lbl);
      return compose(compose(seq->fst, c), s2->snd);
    }
  } else if (TAG(snd) == SEQUENCE_TAG) {
    if (TAG(fst) == FAIL) { return fst; }
    else {
      crcn c = compose(fst, s2->fst);
      return MK_SEQ(c, UNTAG_SEQ(seq)->snd); }
  } else if (TAG(snd) == FAIL) { 
      return TAG(fst) == FAIL ? fst : snd; }
  } else if (TAG(fst) == HAS_2ND_TAG) { 
    snd_tag tag = UNTAG_2ND(fst)->second_tag;
    if (tag == FUN_COERCION_TAG) {
      return compose_fun(fst, snd); }
    else if (tag == REF_COERCION_TAG) {
      ref_crcn r1 = UNTAG_REF(fst); 
      ref_crcn r2 = UNTAG_REF(snd);
      if (read == ID && write == ID) return ID;
      else { 
        crcn c1 = compose(r1->read, r2->read);
        crcn c2 = compose(r2->write, r1->write);
        return MK_REF_COERCION(c1, c2); }
    }
  } else { raise_blame(UNTAG_FAIL(fst)->lbl); }
}
\end{lstlisting}
\caption{The \composecall{} function for normalizing coercions.}
\label{fig:compose-coercion}
\end{figure*}

\end{document}